\documentclass[traditabstract]{aa}
\usepackage{graphicx}
\usepackage{txfonts}

\begin{document}

\title{Massive and Refined. II. The statistical properties of turbulent motions in massive galaxy clusters with high spatial resolution.}

\author{F. Vazza\inst{1,2}, G.Brunetti\inst{2}, C.Gheller\inst{3}, R.Brunino\inst{3}, M.Br\"{u}ggen \inst{1}}

\offprints{Franco Vazza \\ \email{f.vazza@jacobs-university.de}}

\institute{Jacobs University Bremen, Campus Ring 1, 28759, Bremen, Germany
\and  INAF/Istituto di Radioastronomia, via Gobetti 101, I-40129 Bologna,
Italy
\and CINECA, High Performance System Division, Casalecchio di
Reno--Bologna, Italy}

\date{Received / Accepted}

\authorrunning{F. Vazza, G.Brunetti, C.Gheller, R.Brunino, M.Br\"{u}ggen}
\titlerunning{Turbulent motions in massive galaxy clusters.}

\abstract
{We study the properties of chaotic motions in the intra cluster medium using a set of 20 galaxy clusters
simulated with large dynamical range, using the Adaptive Mesh Refinement  code ENZO (e.g. Norman et al.2007). 
The adopted setup allows us to study the spectral and spatial properties of turbulent motions in galaxy clusters
with unprecedented detail, achieving an maximum available Reynolds number of the order of $R_{\mathrm e} \sim 500-1000$ for the largest eddies. The correlations between
the energy of these motions in the Intra Cluster Medium and the dynamical state of the host
systems are studied, and the statistical properties of turbulent motions and their evolution with time support that major merger events are responsible for the injection of the bulk of turbulent kinetic energy
inside cluster. Turbulence is found to account for a $\sim 20-30$ per cent of the thermal energy in merging
clusters, while it accounts for a $\sim 5$ per cent in relaxed clusters.
A comparison of the energies of turbulence and motions in our simulated clusters with present upper-limits in real nearby clusters, recently derived with XMM-Newton (Sanders et al.2010), is provided. When the same spatial scales of turbulent motions are compared, the data from simulations result well within the range presently allowed by observations. Finally, we comment on the possibility that turbulence  may accelerate relativistic particles
leading to the formation of giant radio halos in turbulent (merging) clusters. Based on our 
simulations we confirm previous semi-analytical studies that suggest that the fraction
of turbulent clusters is consistent with that of clusters hosting radio halos.}

\maketitle


\section{Introduction}
\label{sec:intro}

\begin{figure*}
\includegraphics[width=0.95\textwidth]{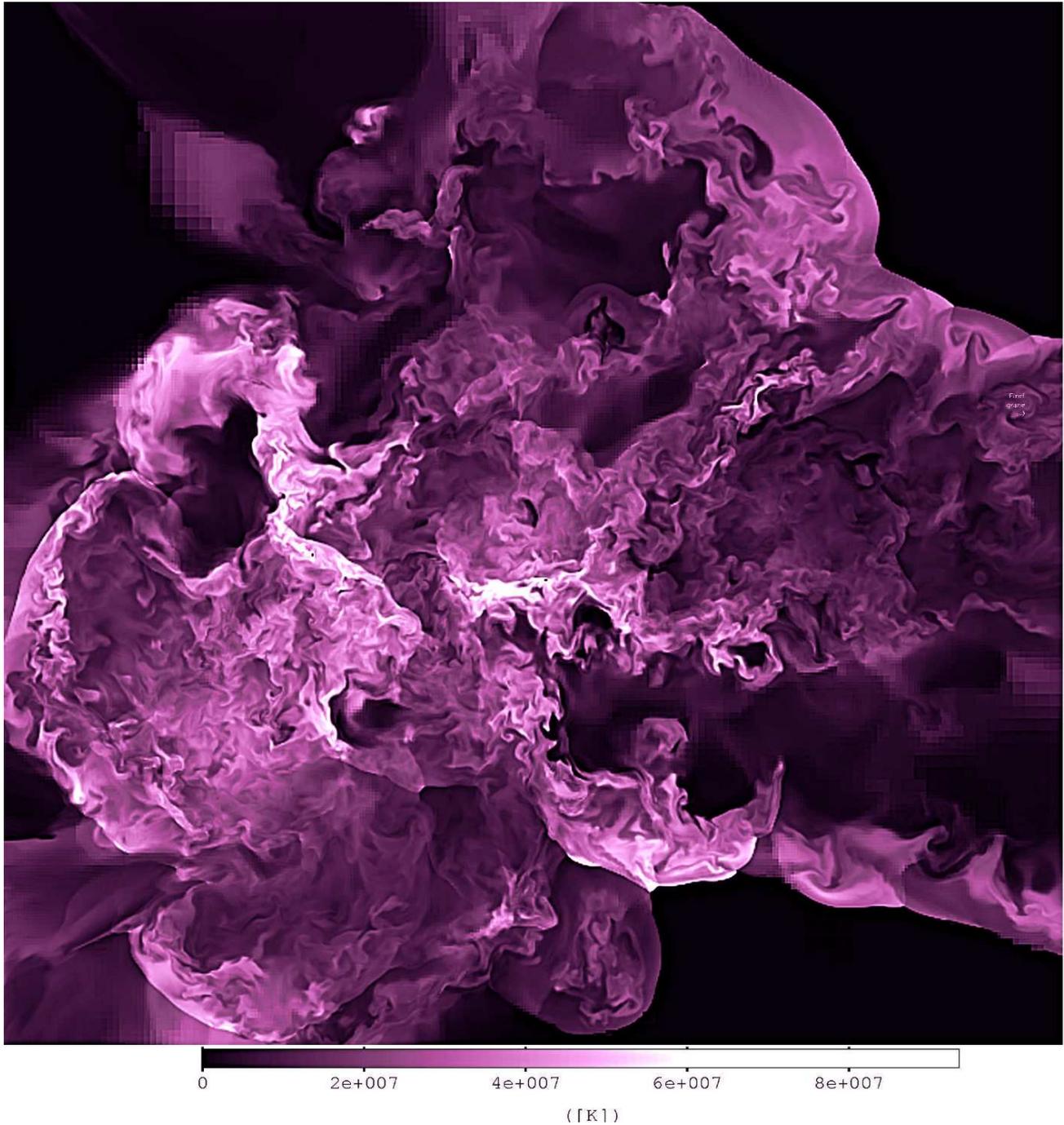}
\caption{2-dimensional slice showing the gas temperature for the innermost
region of  galaxy cluster E1, during its main merger event ($z=0.6$).
The side of the slice is $8.8$~$Mpc ~h^{-1}$ and the depth along the line
of sight is $25$~$kpc~h^{-1}$.}
\label{fig:hell}
\end{figure*}


It is generally believed that turbulent motions are 
generated during the hierarchical assembly of matter in galaxy clusters 
that occurs through mergers and accretion of satellites.

Mergers between galaxy clusters are very energetic events where a
fraction of the kinetic energy of Dark Matter (DM) halos is transferred
into the thermal and kinetic energy. The gas from infalling halos is stripped within
cluster cores due to the action of
ram pressure and several fluid instabilities. This drives
turbulent motions, which may eventually transfer energy from large to smaller
scales (e.g. Sarazin 2002; Cassano \& Brunetti 2005; 
Subramanian et al. 2006; Brunetti \& Lazarian 2007).
Also the sloshing motions of DM cuspy cores that may occur
in the cores of clusters at later stages of merging events 
can drive turbulent motions 
(e.g. Markevitch \& Vikhlinin 2007; Roediger et al.2010).
Even without considering the process of hierarchical formation,
galaxy clusters host several potential sources of turbulent 
motions in the Intra Cluster Medium (ICM).
The activity of AGN that are found in the central cluster
galaxies may excite turbulence in the cores of the hosting
clusters (e.g. Fabian et al. 2003; Rebusco et al.2005; 
Scannapieco \& Br\"uggen 2008), as well as galaxies that
cross the cluster atmosphere (e.g. Deiss et al 1996;
Subramanian et al. 2006).

From the theoretical viewpoint, turbulence may have a deep impact on the physics of the ICM
(e.g. Narayan \& Medvedev 2001; 
Shekochihin et al 2005, 2010; Lazarian 2006; 
Ruszkowsky \& Peng 2010) and on the properties of non-thermal
components in galaxy clusters (e.g. Subramanian et al.
2006; Brunetti \& Lazarian 2007; Brunetti \& Lazarian 2010).
The presence of turbulent gas motions in the ICM is suggested
by measurements of the Faraday Rotation of the polarization
angle of the synchrotron emission from cluster radio galaxies.
These studies show that the magnetic field in the ICM is
tangled on a broad range of spatial scales (e.g. Murgia et al.~2004; 
Vogt \& Ensslin 2005; Bonafede et al.~2010; Govoni et al.~2010; Vacca et al.~2010) suggesting the presence of super--Alfv\'enic 
motions in the medium.
Also independent attempts from X--ray observations of a number of
nearby clusters, based on  
pseudo--pressure maps of cluster cores and on the lack of evidence 
for resonant scattering effects in the X-ray spectra, provided
hints of turbulence in the ICM (e.g. Schuecker et al.~2004; 
Henry et al.~2004; Churazov et al.~2004; Ota et al. 2007).
Important constraints on the fraction of the turbulent and thermal
energy in the cores of clusters are based on the analysis of the broadening
of the lines in the emitted X--ray spectra of cool core clusters
(Churazov et al.2008; Sanders et al.~2010a,b).
In several cases these studies derived interesting upper limits 
to the ratio of turbulent and thermal energy, of the order of $\leq 20$ per cent,
even if the most stringent limits
refer to clusters with very compact cores and are thus sensitive
to (the energy of) motions on a few tens kpc scale.

Nowadays numerical simulations provide a unique way to study the generation and
evolution of velocity fields in the ICM across a wide range of scales.
Early Eulerian numerical simulations of merging clusters (e.g.,
Roettiger, Stone \& Burns 1999; Norman \& Bryan 1999; Ricker \& Sarazin 2001)
obtained the first reliable representation of the way in which 
turbulence is injected into the ICM by merger events. 
More recent works, performing high-resolution
Lagrangian (Dolag et al.~2005, Vazza et al.~2006; Valdarnini 2010) 
and Eulerian re-simulations of galaxy clusters  extracted from large 
cosmological volumes  found that a sizable
amount of pressure support (i.e. $\sim 10-30$ percent of the total 
pressure inside $0.5 R_{\mathrm vir}$ {\footnote{In this paper, we adopt the customary definition of the virial radius, $R_{\mathrm vir}$, as the radius enclosing
a mean total (gas+DM) overdensity of $\approx 109$ times the critical density of the Universe (e.g. Eke 1998). }}) in the ICM can be due to 
chaotic motions, provided that the kinematic viscosity in the 
innermost cluster region is negligible for scales $\geq 10$~kpc (Nagai et al.~2007; Iapichino \& Niemeyer 2008; 
Lau et al.~2009; Vazza et al.~2009; Paul et 2010; Burns, Skillman \& O'Shea 2010).

\noindent
The Adaptive Mesh Refinement technique (AMR) is an
optimal approach to study the fluid-dynamics of the evolving
ICM with high spatial resolution, and within a 
fully cosmological framework.
This technique represents an efficient way to overcome the problem of having 
a too coarse spatial resolution
in the central region of collapsed objects, typical of fixed mesh Eulerian simulations 
(e.g. Berger \& Colella 1989).
The proper triggering of the mesh-refinement criteria in AMR
simulations allows for reaching  
a peak resolutions comparable to the Lagrangian (Smoothed Particles 
Hydrodynamics, SPH) simulations in the innermost
cluster regions (e.g., $\sim 10-20$ ~kpc), while preserving the 
shock-capturing nature of the algorithm (e.g. Teyssier 2002).

\bigskip
Recent works employing cosmological simulations have shown that the 
use of mesh refinement criteria anchored to the properties of  
the 3--D velocity field in the ICM enhances the numerical 
description of chaotic motions (e.g. Iapichino \& Niemeyer 2008;
Vazza et al.~2009; Vazza, Gheller \& Brunetti 2010).
Furthermore, the adoption of sub-grid modeling to treat the dynamical role 
played by unresolved gas motions helps to model this issue 
more properly, although the importance of these motions
in the context of galaxy clusters is not totally
clear (e.g. Scannapieco \& Br\"{u}ggen 2008; Maier et al.~2009).

In Vazza et al. (2010, hereafter paper I) we presented a sample of 
20 massive galaxy clusters re-simulated with the code ENZO (Norman et al.2007) and employgin an AMR strategy designed to increase the spatial resolution 
in shocks and turbulent features in the ICM up to 
distances of $\sim 2-3$ virial radii from cluster centers {\footnote{A public archive of this
data-sample is accessible at {\it http://data.cineca.it}}}.
The general properties of the thermal gas distribution and of 
shock waves in the sample were discussed in Paper I, in relation
to the dynamical history of each cluster object. 

In the present paper, we focus on the characterization of the chaotic
motions of the ICM in these  simulated clusters.

\noindent
Compared to previous works on the same issue, our approach combines a large
statistics (which allows us to derive viable correlations between turbulent and dynamical
features) with high spatial resolution (which enables to follow the details of the gas dynamics
and their spectral properties achieving a good scale separation).

The paper is organized as follows: in Sec.\ref{sec:methods} we present the re-simulation
technique adopted to produce the sample of clusters, and its general properties. In Sec.\ref{sec:results}
we discuss the characterization of turbulent gas motions in the sample, focusing on the
radial distribution of turbulent motions 
within clusters, on the spectral features of the 
ICM velocity field across the sample and on the scaling
relations found between turbulent energy and the virial parameters of the host clusters. We explore in detail
 the possible connection between the statistical features of large-scale turbulent motions in
our cluster sample and the observed statistics of radio halos emission in Sec.\ref{subsec:halos}. 
Finally, we summarize our conclusions in Sec.\ref{sec:conclusions}.

\begin{figure} 
\begin{center}
\includegraphics[width=0.48\textwidth]{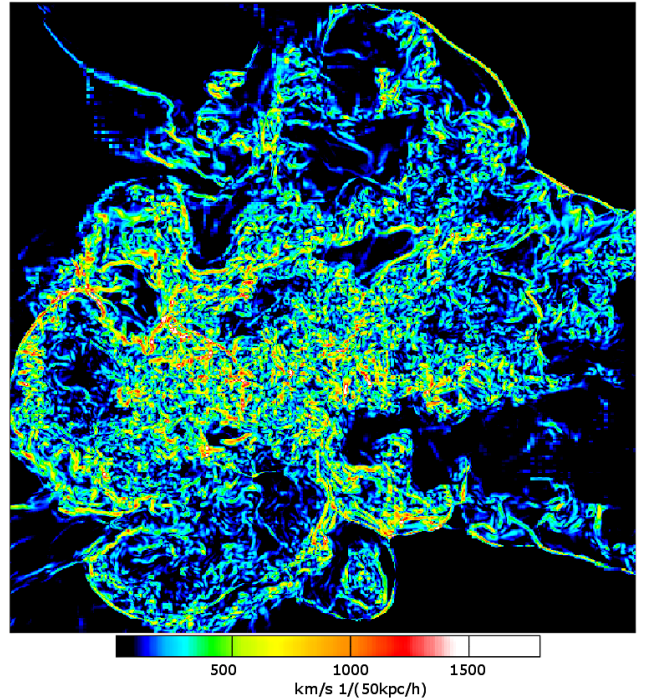}
\caption{Slice showing the curl of velocity for the same region of
Fig.\ref{fig:hell}. }
\label{fig:curl}
\end{center}
\end{figure}

\begin{figure*}
\includegraphics[width=0.95\textwidth]{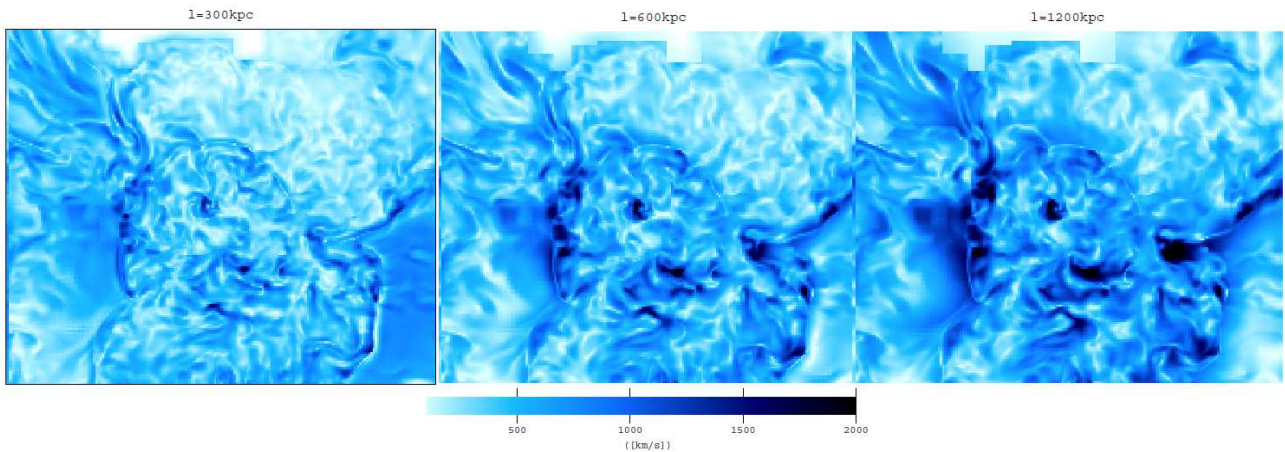}
\caption{Maps of the absolute value of the chaotic velocity field for three
choices of the coherence scale for $V_{\mathrm l}$ ($l_{\mathrm MAX}=300$kpc, $=600$kpc and $=1200$kpc).
The side of the slices is $4$~$Mpc ~h^{-1}$ and the depth along the line
of sight is $25$~$kpc ~h^{-1}$.}
\label{fig:filter}
\end{figure*}

\section{Clusters simulations.}
\label{sec:methods}

Computations presented in this work were performed using the 
ENZO 1.5 code developed by the Laboratory for Computational
 Astrophysics at the University of California in San Diego 
(http://lca.ucsd.edu).

The details for the numerical setup are described in Paper I.
In summary, cosmological initial conditions were produced with 
nested grid/DM particle distributions of increasing mass 
resolution to achieve a high DM mass resolution 
($m_{\mathrm DM}=6.76 \cdot 10^{8} M_{\sun}$)
in the region of formation of each cluster, of volume $\approx (5-6 R_{\mathrm vir})^{3}$, where $R_{\mathrm vir}$ is the final virial radius for each clusters.

The clusters were extracted from several  boxes sampling a total cubic cosmic volume of 
$L_{\mathrm box} \approx 440$~~$Mpc ~h^{-1}$.

We assumed a ``concordance'' $\Lambda$CDM cosmology with
$\Omega_0 = 1.0$, $\Omega_{\mathrm BM} = 0.0441$, $\Omega_{\mathrm DM} =
0.2139$, $\Omega_{\Lambda} = 0.742$, Hubble parameter ${\it h} = 0.72$ and
a normalization for the primordial density power
spectrum $\sigma_{\mathrm 8} = 0.8$.

Our runs neglect radiative cooling, star formation and feedback from AGNs, while
re-ionization is treated with a simplified Haardt \& Madau (1996) re-ionization model (see Appendix of Paper I for details).

The physical modeling of re-ionization in our runs is mainly motivated by the requirement of reproducing 
realistic shock waves, since
re-ionization is expected  to play a sizable role in setting the strength of outer accretion shocks (e.g. Vazza, Brunetti \& Gheller 2009).

The clusters in the sample have total masses larger than $6 \cdot 10^{14}M_{\sun}$,
12 of them having a total mass above $>10^{15}M_{\sun}$. 

Grouping the clusters according to their dynamical
state is useful to detect statistical properties 
which vary inside the cluster sample.  In Paper I, we outlined our procedure to distinguish between 
{\it "relaxing"}, {\it "merging"} and {\it "post merger"} systems of our sample, focusing
on the total matter accretion history of each halo.
According to our definition, this sample contains
10 post-merger systems (i.e. clusters with a merger with a mass ratio larger than $1/3$ for $z \leq 1$), 
6 merging clusters at $z=0$ and 4 relaxing clusters (i.e. systems without evidence of past or ongoing major merger for $z<1$).

The refinement strategy inside the $\approx (5-6 R_{\mathrm vir})^{3}$ volume
(hereafter the ``AMR region'') was tuned to increase the grid resolution
based on gas/DM over-density, $\delta \rho_{gas,DM}$ and/or velocity jumps.
In this scheme, a normalized 1--D velocity jump across 3 close cells in the scan direction (at a given refinement level) is computed
as $\delta_{\mathrm v} \equiv |\Delta v|/|v_{\mathrm min}|$, where $|v_{\mathrm min}|$ is the minimum velocity among the 3 cells;  
the threshold values used to trigger the mesh refinement were $\delta \rho_{\mathrm gas,DM}/\rho_{\mathrm gas,DM}>3$ for gas/DM over-densities and $\delta_{\mathrm v}>3$ for the 1--D velocity jump.
For tests about the numerical convergence of this method and for a comparison with the more standard refinement based on gas/DM over-density, we address the readers to our previous works (e.g. Vazza et al.2009; Vazza ~2010).

The adoption of this refinement criteria was necessary 
to avoid the spurious suppression of turbulent eddies moving from
dense to less dense regions. This is a major drawback of Lagrangian approaches (e.g. Smoothed
Particles Hydrodynamics, e.g. Dolag et al.~2008 for a review) and of grid methods for which 
AMR is triggered by over-density criteria only (e.g. Bryan \& Norman ~1998). 
In general, the AMR region around each of our cluster at $z=0$ is always sampled with at least ${\it N} \sim 500^{3}$ cells at the highest available resolution.


When this extra refinement criterion is adopted, turbulent eddies can evolve 
with the maximum possible dynamical range, and the decay of turbulence  is not
artificially damped by numerical under-sampling.
 A first order estimate of the Reynolds number available to the largest possible eddies contained within the AMR region yields 
a value of the order of ${\it R_{\mathrm e}} \sim N^{4/3} \sim 4000$ ~(e.g. Kritsuk, Norman \& Padoan 2006). However, based on the velocity 
power spectra measured for the 3--D velocity field in our clusters (see Sec.\ref{subsec:spectra}), the maximum Reynolds number
which achieved in our simulated ICM is the range of ${\it R_{\mathrm e}} \sim N^{4/3} \sim 500-1000$.
In any case, a rather large hierarchy of chaotic "eddies" of decreasing size can develop within the AMR region and evolve towards
the scale of the numerical dissipation, which is a few times the minimum cell size {\footnote{For a discussion
of the effect of PPM artificial dissipation at the smallest scales in the simulation see however the discussion
in Sec.\ref{subsec:spectra}.}}.

To illustrate the vast amount of spatial details provided by the method employed here, in Fig.\ref{fig:hell} we present
an example of the pattern of temperature fluctuations in a multiple mergers event in one of our
forming galaxy clusters (E1), at the time it
assembled the bulk of its total mass ($z \sim 0.5-0.6$). The side of the slice is $\approx
8.8$~ ~$Mpc ~h^{-1}$ and the width is $25$~~$kpc ~h^{-1}$.
The rise of fluid instabilities and vorticity at oblique shocks is evident behind the wake of accretion shocks at the cluster periphery, thanks to the sampling at the peak resolution ($25$~$kpc ~h^{-1}$) even at $4-6$~$Mpc ~h^{-1}$ from the cluster center.
A patchy multi-temperature ICM is observed at this stage of the merger, and a multi-scale distribution of 
eddies is generated in the core cluster region by the large $\sim 1000-2000$~ ~$km ~s^{-1}$ relative 
velocities of the colliding clumps. 

The small-scale chaotic components of the ICM flow are well highlighted by detecting the rotational part of the 3--D 
velocity field (e.g. Ryu et al.2008; Zhu, Feng \& Fang 2010; Paul et al.2010). 
Figure \ref{fig:curl} shows the map of $|\nabla \times \vec{v}|$ for the same region of Fig.\ref{fig:hell}.
 We compute the vorticity at the scale of the cells with a simple first-order finite difference method, in which the
difference is computed across a baseline scale of $l_{\mathrm curl} = 50$~$kpc ~h^{-1}$ in the three directions across each cell.
We find that accretion and merger shocks inject a vorticity of the order of $\sim 1000-1500$~$km ~s^{-1}$ on the scale of $l_{\mathrm curl}$ at the location of the outer accretion shocks, while
shears along the stripped infalling sub-clumps
(e.g. the cold stream of gas entering on the lower right corner in Fig.\ref{fig:hell}-\ref{fig:curl}) inject small scale vorticity at a lower rate, $\sim 300-600$~$km ~s^{-1}$, but over larger volumes.

\section{Results}
\label{sec:results}

\begin{figure}
\includegraphics[width=0.45\textwidth]{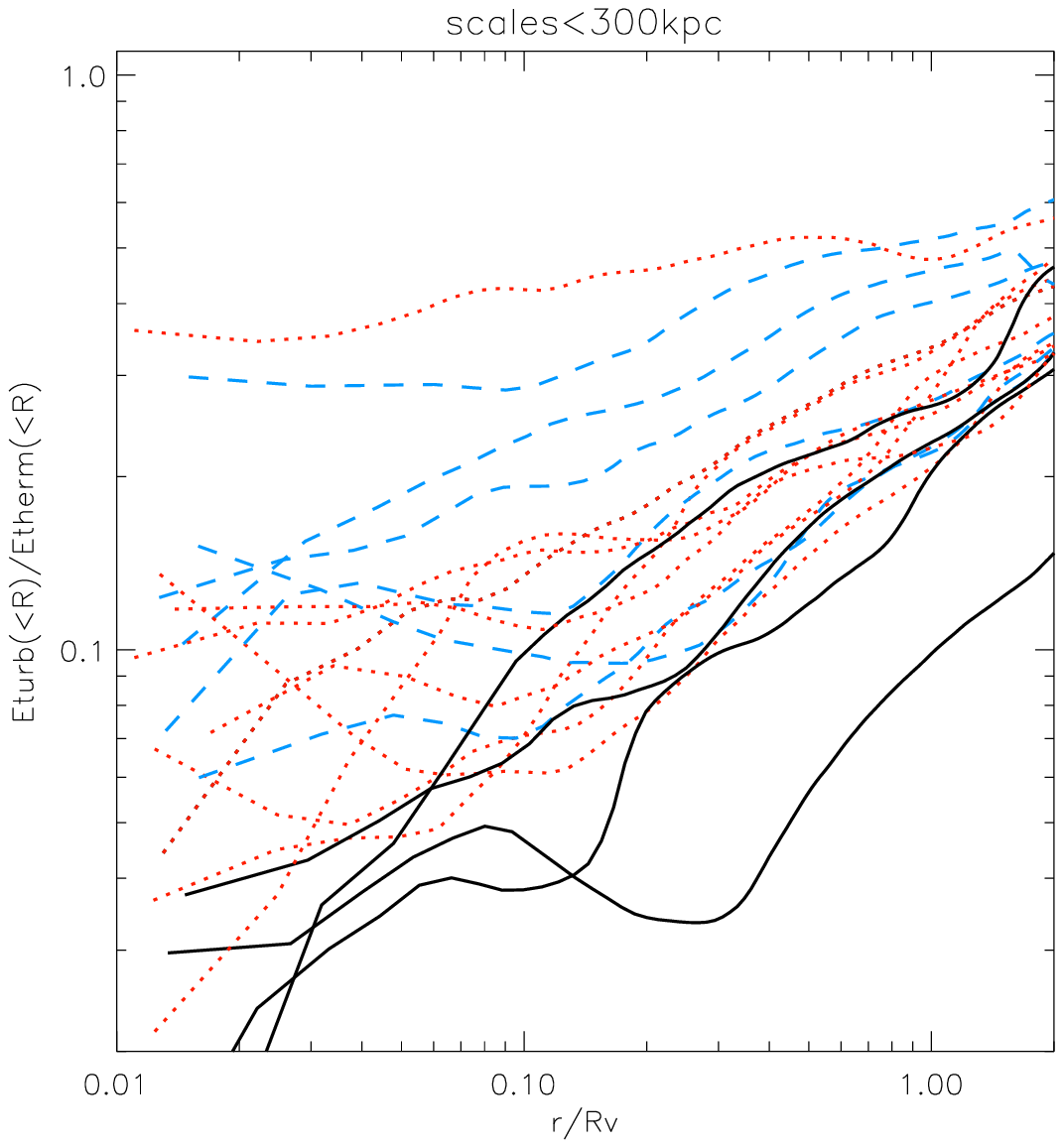}
\includegraphics[width=0.45\textwidth]{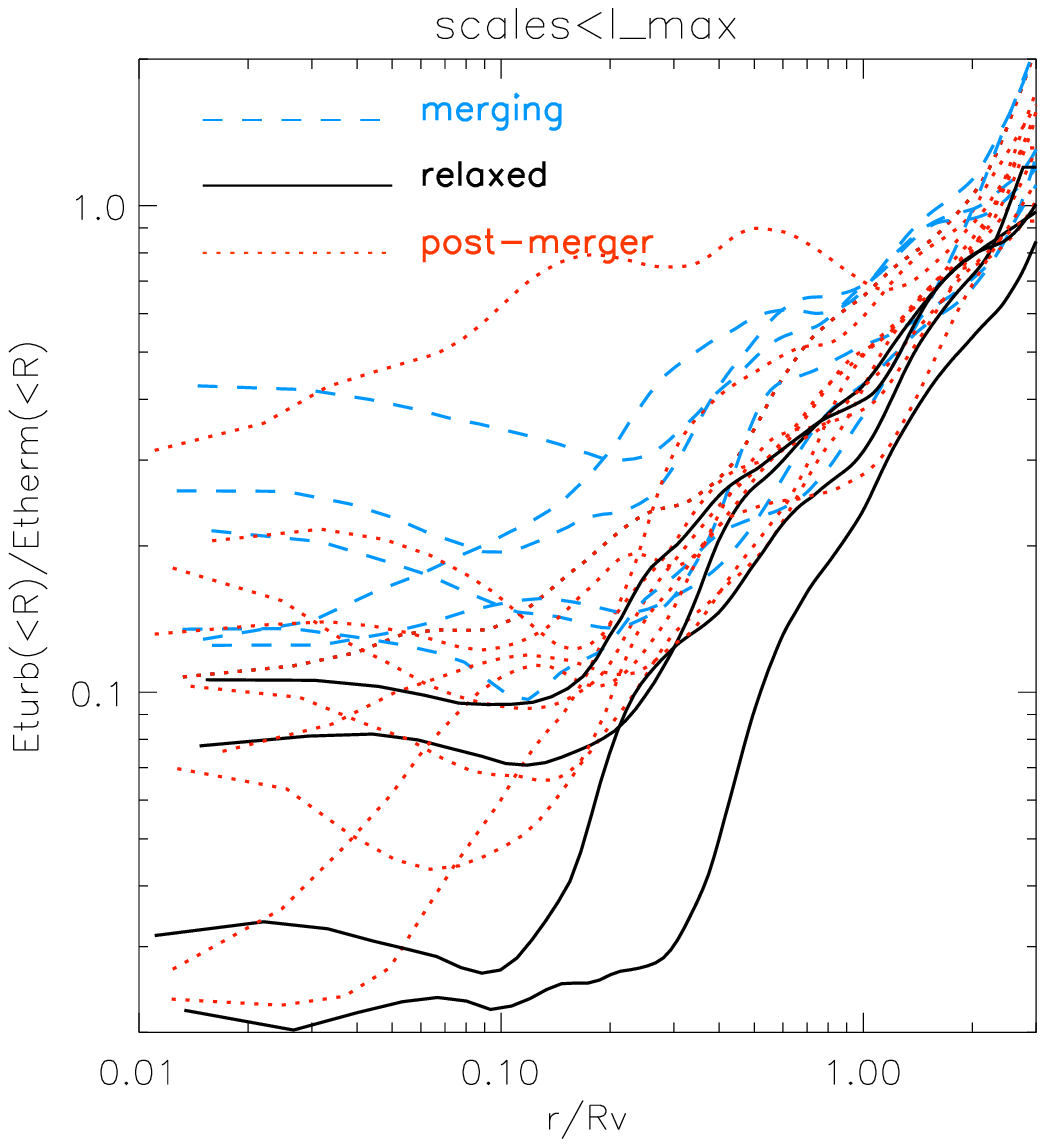}
\caption{Profiles of the turbulent to total energy ratio for $l_{\mathrm MAX}=300$kpc (top panel)
and for $l_{\mathrm MAX}=2 \pi/k_{\mathrm MAX}$ (lower panel), at $z=0$. The different colors refer to clusters with a different
dynamical state, as explained in Sec.\ref{sec:methods}.}
\label{fig:prof_ratio}

\end{figure}

\subsection{The decomposition of "bulk" and "turbulent" motions.}
\label{subsec:turbo_method}

In order to characterize turbulent velocity 
fields in the complex environment of galaxy clusters,
it is necessary to extract the velocity 
fluctuations from the complex 3--D distribution of velocities. 
This is not a trivial task and a number of different
strategies have been adopted in the recent past to 
characterize the "bulk" and "turbulent" component of the velocity field.
This has been done for instance by taking spherical shell averages (e.g. Bryan \& Norman 1999;
Iapichino \& Niemeyer 2008; Lau et al.2009; Burns et al.2010), by mapping in 3--D 
the velocity field on a resolution coarser than the maximum resolution
of the simulation (e.g. Dolag et al.2005; Vazza et al.2006; Vazza et
al.2009; Valdarnini 2010), or by using sub-grid modeling (e.g. Maier et al.2009).

In this work we adopt two different methods to filter the large-scale component of the velocity field of the ICM, in order to account
for the uncertainty inherent in the particular representation of the 
turbulent ICM. 

In the first method (hereafter the "$k_{\mathrm MAX}$" method) we calculate the power spectrum of the
3--D velocity field of the ICM and identified the spatial frequency containing most
of the kinetic energy in the flow. According to the simplest view of turbulent fluids,
this maximum scale $k_{\mathrm MAX}$  marks the beginning of the energy cascade
of turbulent eddies down to the dissipative scale available to the simulation (which is of the
order of a few cells at the highest available resolution) and would also
represent the scale of the maximum Reynolds number of the flow.

Following Vazza et al.(2009) and Vazza, Gheller \& Brunetti (2010) we
measured the power spectrum of the velocity field of the simulated ICM, $E(k)$, defined as:

\begin{equation}
E({\vec {k}}) = {1\over 2} |{\vec{ \widehat{v}(k)}}|^{2},
\label{eq:pek}
\end{equation}

where ${\vec{\widehat{v (k)}}}$ is the Fourier transform of the the 3--D velocity field (${\vec {v}}=[v_{x},v_{y},v_{z}]$) defined as :

\begin{equation}
{\vec {\widehat{v(k)}}}=
\frac{1}{(2\pi)^3}\int_{V}{\vec v (x)}e^{-2\pi i \,{\vec{k \cdot x}}}d^{3}x.
\end{equation}

The 3--D velocity is measured in the center of (total) mass reference
frame, and the power spectrum is calculated with a
standard FFT algorithm on the data at the highest available resolution, with the addition of 
a zero-padding technique to deal with the non-periodicity
of the considered volume and of a Gaussian apodization function to
avoid the generation of spurious frequencies at the box edges (see Vazza, Gheller \& Brunetti 2010 and Valdarnini 2010 for a discussion).

The power spectrum is measured within a cubic region with
side $\sim 4R_{\mathrm vir}$ centered on each cluster,
and the scale $k_{\mathrm MAX}$ is set as the scale of the maximum value of $k \cdot E(k)$. 
Then for every cluster we filter out
the velocity component $\vec {\widehat v(k)}$ with $k \leq k_{\mathrm MAX}$, and consider
 "turbulent" only the inverse transform in the real space of the filtered
$\vec  {\widehat v(k>k_{\mathrm MAX})}$.
For the clusters of the sample $k_{\mathrm MAX}$ is found to correspond to a maximum spatial scale of $l_{\mathrm MAX} \sim 0.75 - 1.5 R_{\mathrm vir}$ .
Since laminar bulk motions of coherence scales smaller than
$l_{\mathrm MAX}$ are not filtered out from this approach, we expect that this method 
provides a {\it higher limit} to
the content of the kinetic turbulent energy in the ICM.

In the second method, following previous works 
on the same topic (e.g. Dolag et al.~2005; Vazza et al.~2006; Vazza et al.~2009; Vazza, Gheller \& Brunetti 2010)
 we mapped the 3--D velocity field using the fixed spatial scale of $l_{\mathrm MAX}=300kpc$ as a filter
to highlight the ''turbulent'' small scale features of the ICM. 
The filtering procedure works is applied directly in the real space, by interpolating the 
original 3--D velocity field with a Triangular Shape Cloud kernel (e.g. Hockney \& Eastwood 1981)  of width $l_{\mathrm MAX}=300$~kpc
to map the local mean field, $V_{\mathrm l}$, and to detect  turbulent fluctuations on small scales.
 This is motivated by early studies based on SPH (Dolag et al.2005) which
suggested that the typical size of gas/DM clumps crossing the
virial volume of galaxy clusters with  $M > 10^{14} M_{\sun}/h$  at  $z \sim 0$
is $<200-300$~kpc, and that therefore this filtering scale is suitable to 
detect velocity fluctuations on scales smaller than the scale of typical laminar
infall motion driven by the accreted satellites.
However, in the present work we consider large clusters (by a factor $\sim 10$
in total mass and by a factor $\sim 3$ in $R_{\mathrm vir}$) compared to previous works. 
Therefore, a larger physical scale for the injection of
turbulent motions by accretion of sub-clumps may by expected and therefore 
this may generally yield a {\it lower limit}
on the estimate of turbulent energy within our simulated clusters.  

As an example of the effects of adopting different scales to estimate the turbulent
energy budget, in Fig.\ref{fig:filter}  we show the progressive change introduced by the filtering length used 
to compute $V_{\mathrm l}$ on a cluster simulation.
The change of the chaotic field with filtering length {\it l}
is more evident in the outermost regions, which are characterized also by bulk motions
with coherence scales of the order of 
$\sim 300-500$~kpc (see the evolution of the clumps located at the right of the
cluster center in Fig.\ref{fig:filter}). The pattern of motions
in the central cluster regions are found to be tangled at scales much smaller than a Mpc, therefore
adopting $300$~kpc as a filtering length (or more) does not produce significant differences
in the reconstructed turbulent velocity field.

In the following, we will show
that the two filtering techniques produce statistically consistent results
when applied to the whole cluster sample, and therefore that the statistical
features associated with the turbulent ICM are rather independent of the
particular filtering method adopted here.

\begin{figure}
\includegraphics[width=0.48\textwidth,height=0.43\textwidth]{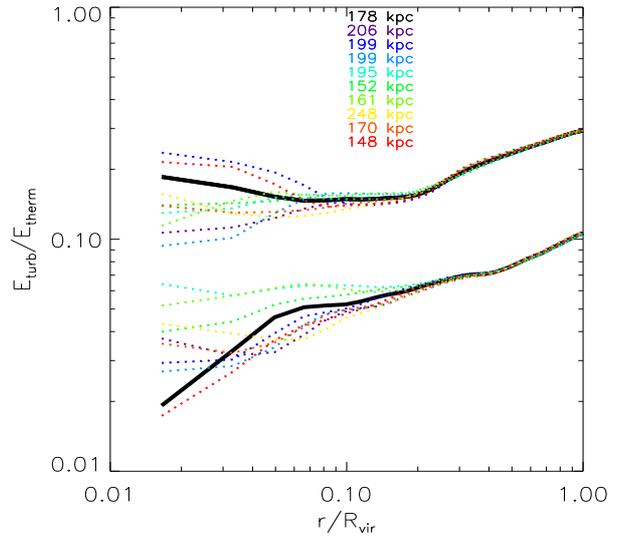}
\caption{Average profiles of the cumulative turbulent to total energy ratio at $z=0$, for 9 random displacements from the putative center of mass of the cluster (solid black profiles). The top curves are for a post-merger cluster, the bottom curves are for a relaxed cluster. The different colors refer to different displacements from the center (the corresponding absolute values are reported in the panel).}  
\label{fig:prof_test}
\end{figure}

\begin{figure}
\includegraphics[width=0.48\textwidth,height=0.48\textwidth]{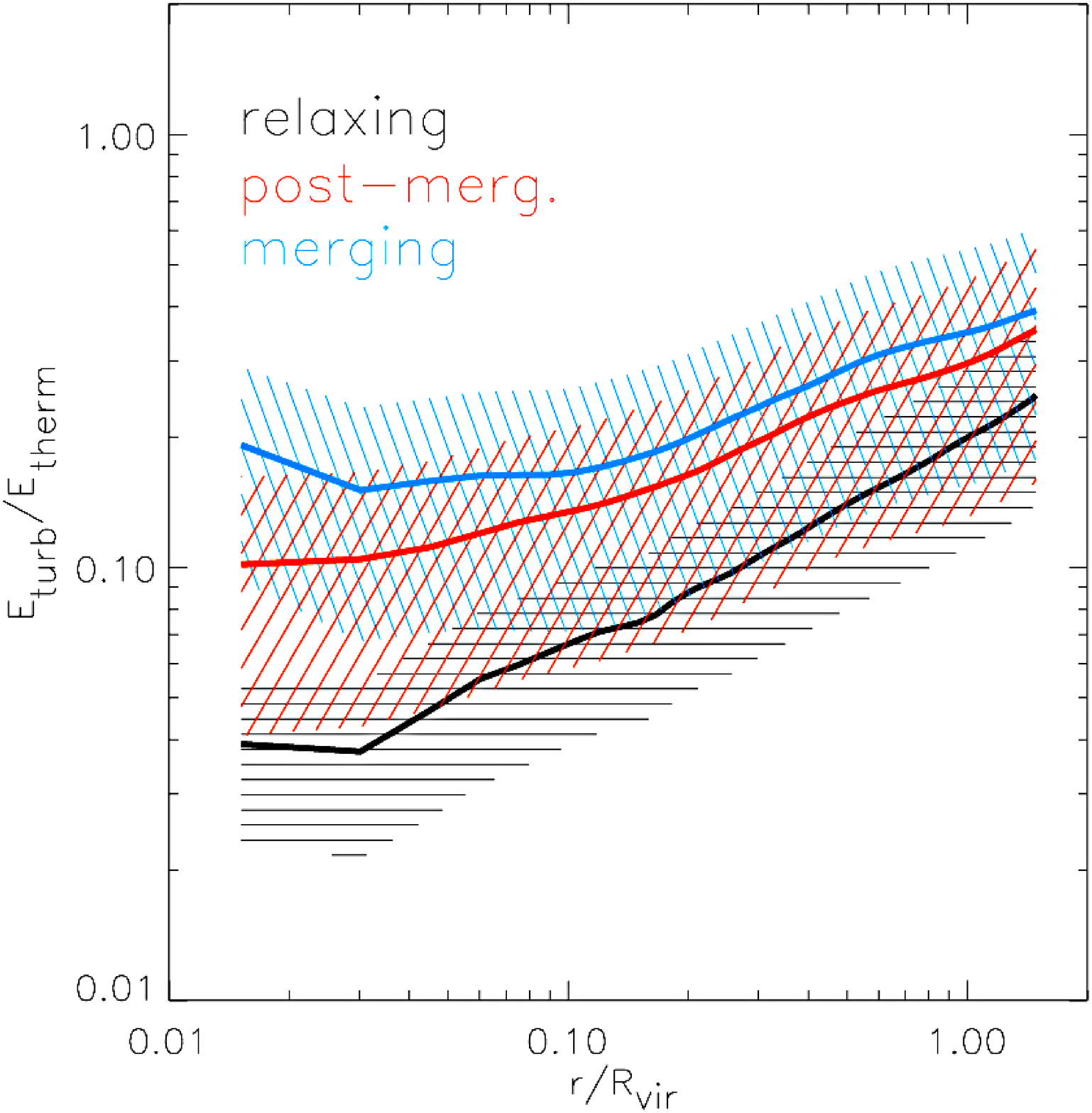}
\includegraphics[width=0.48\textwidth,height=0.48\textwidth]{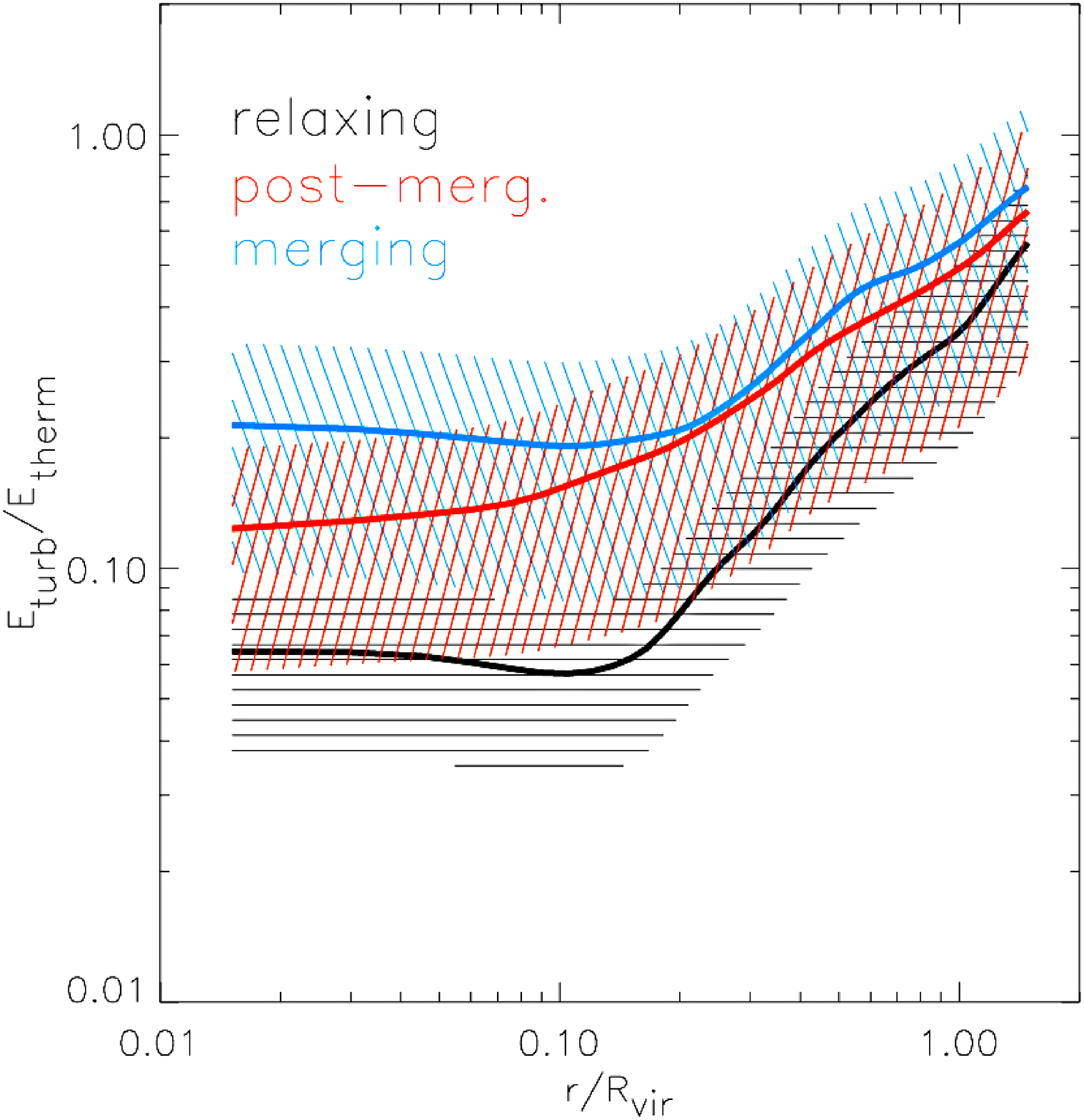}
\caption{Average profiles of the turbulent to total energy ratio at $z=0$, for the 3 dynamical classes of clusters and adopting  $l_{\mathrm MAX}=300$kpc for the filtering of
the local velocity field (top) or $l_{\mathrm MAX}=2 \pi/k_{\mathrm MAX}$ (bottom).} 
\label{fig:prof_classes}
\end{figure}


\subsection{Radial distribution of the turbulent energy in the ICM.}
\label{subsec:radial}

The ratio between the turbulent energy of the ICM,  
$E_{\mathrm turb}=\rho v_{t}^{2}/2$ (where $v_{t}$ is the "turbulent"  velocity field, ${\vec {v_{t}}}={\vec {v}}-{\vec {V_{\mathrm l}}}$) and the total thermal energy, $E_{\mathrm therm}=3/2 k_{B} T \rho/(\mu m_{p})$ (where
$T$ is the gas temperature, $k_{B}$ is the Boltzmann's constant, $\mu$ is the mean molecular
mass and $m_{p}$ is the proton mass), is a simple and important proxy of the
importance played by turbulent motions in clusters dynamics.

\bigskip

Figure \ref{fig:prof_ratio} shows the behavior of the cumulative ratio $E_{\mathrm turb}/E_{\mathrm therm}$
inside a given radius for all the clusters in the sample at $z=0$; for clarity, we
show with different colors the three dynamical classes discussed in Sec.\ref{sec:methods},
and normalize all radii to the $R_{\mathrm vir}$ of each cluster.
\noindent
All systems show a profile that increases with radius, with the smallest values of 
$E_{\mathrm turb}/E_{\mathrm therm}$ (in the range $<10-20$ per cent for most of clusters) inside $0.1 R_{\mathrm vir}$.
This ratio reaches larger values at $R_{\mathrm vir}$, $E_{\mathrm turb}/E_{\mathrm therm} \sim 0.5-0.6$.
In the case of relaxed systems the turbulent energy stored inside the core region
of the clusters can be as small as $\sim 5$ per cent, while in merging or post-merger systems
it can be as large as $\sim 20-30$ per cent. 
Inside $0.1R_{\mathrm vir}$, two systems have $E_{\mathrm turb}/E_{\mathrm therm} >0.3$ with both methods, while inside $0.5R_{\mathrm vir}$  4 (7) clusters have $E_{\mathrm turb}/E_{\mathrm therm} >0.3$ if the 
filtering is done using $l_{\mathrm MAX}=300kpc$ ($l_{\mathrm MAX}=2 \pi/k_{\mathrm MAX}$).

 Some caution should be taken in analyzing the radial distribution of turbulence of Fig.\ref{fig:prof_ratio} in the innermost cluster regions ($r<0.1 R_{\mathrm vir}$),
since the characterization of the center of mass of some systems may be subject to uncertainties, particularly in the case of major merger events. In these cases, asymmetries in the matter distribution and large ($\sim 100$~kpc) displacements between the centers of gas matter and DM  can cause uncertainties
in the characterization of the cluster centers. In this respect, to test the stability of our results, we measure the radial
distribution of $E_{\mathrm turb}/E_{\mathrm therm}$ ($l_{\mathrm MAX}=300~kpc$) assuming 9 different random centers extracted inside a sphere of $\approx 250$~kpc around the peak of total mass, for a post-merger cluster and for a relaxed one (Fig.\ref{fig:prof_test}). 
In both system, the differences in the estimated turbulent energy ratio inside $r<0.1 R_{\mathrm vir}$ can be as large as a factor $\sim 2-3$; however even in the case of the perturbed system a good convergence in the estimated $E_{\mathrm turb}/E_{\mathrm therm}$ is achieved for $r \sim 0.1 R_{\mathrm vir}$, even for extreme values of the assumed displacement from the ``real'' cluster centers. With this caveat in mind, in what follows we will mainly focus on the integrated values of $E_{\mathrm turb}/E_{\mathrm therm}$ on radii larger than $r>0.1 R_{\mathrm vir}$, observing that in general the differences between the dynamical classes of clusters are larger than the uncertainty associated to the exact location of the cluster center.

\bigskip

Overall, when compared outside of the innermost region, the profiles of turbulent energies in Smoothed Particles Hydrodynamics simulations of galaxy clusters (e.g. Dolag et al.~2005; Valdarnini~2010) and in AMR grid simulations as here (see also Iapichino \& Niemeyer~2008; Vazza et al.~2009) present a good similarity. However, small sistematic differences are present, also due 
to the different  ways in which gas matter from infalling clumps is stripped in the two methods (e.g. Agertz et al.~2007) and due to the different
stratification of gas entropy within the inner cluster atmosphere,  
which in turn affects the stability to convective motions in the ICM (e.g. Wadsley et al.2008; Mitchell et al.2008; Springel 2010; Vazza 2010).

 \bigskip

\begin{figure*}
\begin{center}
\includegraphics[width=0.975\textwidth,height=0.38\textwidth]{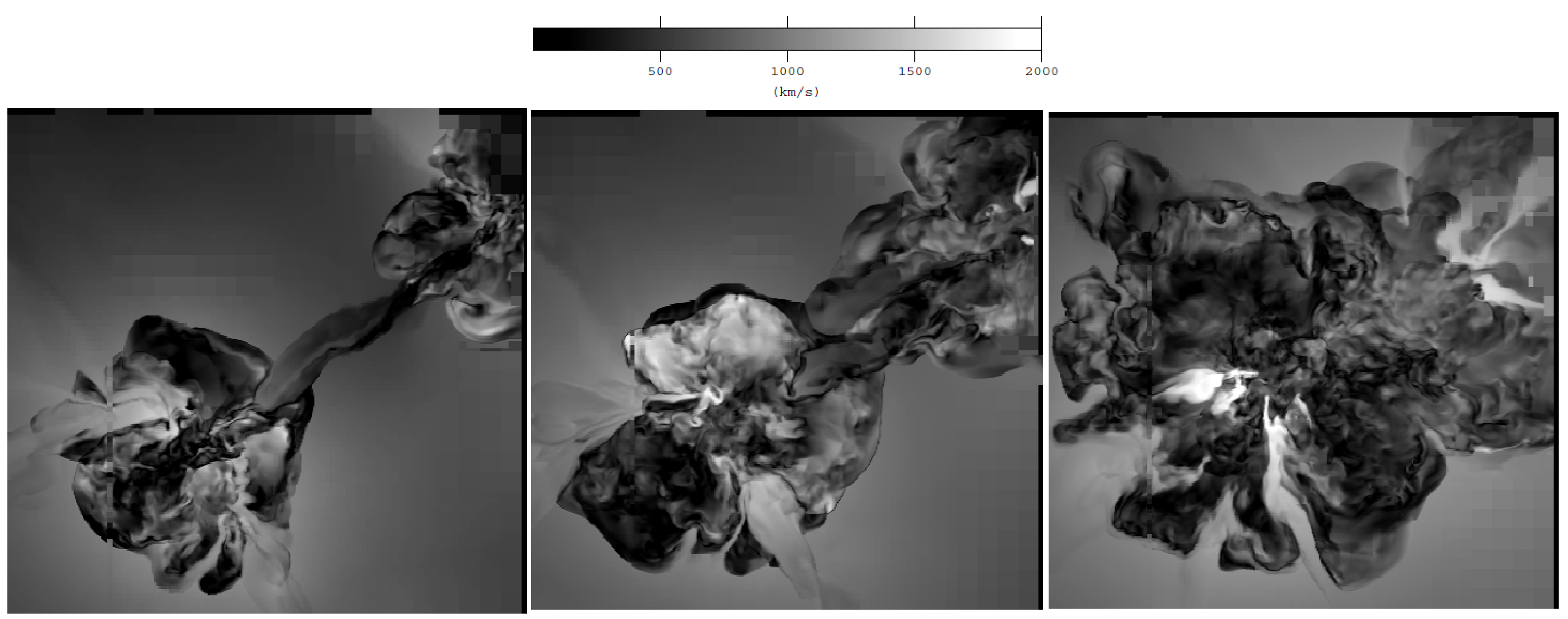}
\includegraphics[width=0.975\textwidth,height=0.38\textwidth]{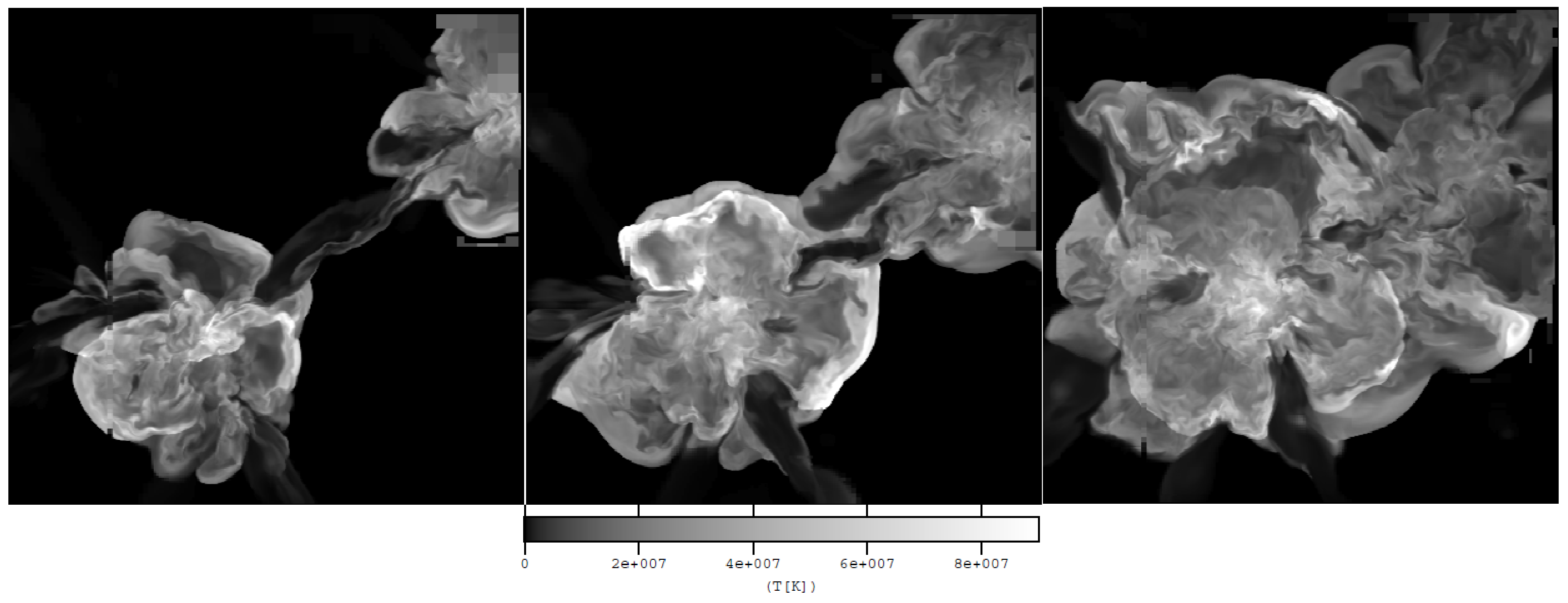}
\caption{Time evolution for a slice cut through the center of the major merger
cluster E26 (from left to right, images are taken at $z=0.79$, $z=0.59$ and $z=0.28$). The bottom
panels show the gas temperature, the top ones show the velocity module. The
side of each slice is of $8$~$Mpc ~h^{-1}$.}
\label{fig:turbo_mov}
\end{center}
\end{figure*}

In order to describe more general features of the three dynamical classes of clusters, we computed the average profiles of all the cluster within each subsample, as shown in Fig.\ref{fig:prof_classes}.
Merging, post-merger and relaxing systems define a "sequence",  with the latter being less turbulent within the virial volume. 
The larger values of $E_{\mathrm turb}/E_{\mathrm therm}$ in merging systems 
suggest that the generation of turbulence starts before the close 
encounter of cluster cores.
In Fig.\ref{fig:turbo_mov} we show the evolution of gas temperature and total gas velocity for a slice
of depth $25$~$kpc ~h^{-1}$ through the merging axis of the major merger cluster E26, for the epochs of
$z=0.79,0.59$ and $0.28$ (the closest cores encounters happens at $z \sim 0.3$). 
The actual major merger is anticipated by a number of minor mergers of gas/DM matter flowing along the massive big filament which connects
the two clusters, triggering shocks and chaotic motions in the cluster outskirts.
The virialization of the outer ICM is much less efficient than in the center, and the temperature is lower compared to the innermost ICM, therefore it is more easy to drive transonic turbulent motions which eventually sink towards the center
of the post-merger clusters. 

\begin{figure*}
\includegraphics[width=0.95\textwidth]{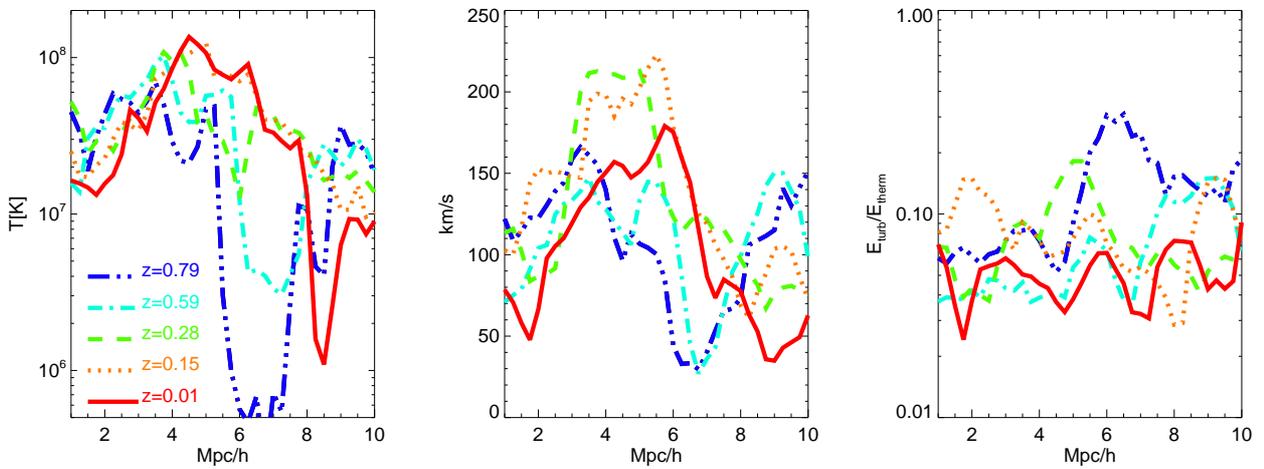}
\caption{Average gas temperature (left), average turbulent velocity (center) and average energy ratio $E_{\mathrm turb}/E_{\mathrm therm}$ (right) along the axis of the major merger of cluster E26, for a cylinder of radius $\sim 300kpc$. The different lines and colors are referred to different redshits.}
\label{fig:cylinder_z}
\end{figure*}

We analyzed the merger sequence of this system by computing the average values of gas temperature, gas turbulent velocity  and $E_{\mathrm turb}/E_{\mathrm therm}$ 
in a cylinder of radius $\approx 300~kpc$ along the axis of merger (Fig.\ref{fig:cylinder_z}).
Even if at the beginning of the merger the turbulent velocity can be fairly small ($\sim 50-100 ~km~s^{-1}$), the energy ratio reaches large values ($E_{\mathrm turb}/E_{\mathrm therm} \sim 0.4$ at $z \approx 0.8$) due to the presence of cold gas falling from the compenetrating outskirts of the two clusters. 
During the later stages of the collision, the heating of merger shock waves increases the temperature of the ICM and decreases the ratio $E_{\mathrm turb}/E_{\mathrm therm}$ , even if the turbulent velocity field experiences an overall increase ($ \sim 200 ~km~s^{-1}$ for $z<0.3$). 
This behavior highlights the important fact that the ratio between turbulent energy and thermal energy does not always mirror a change
in the {\it absolute} turbulent energy/velocity, especially for those regions (or stages) in which the ICM is in a highly un-virialized state.

A qualitatively similar complex picture of galaxy clusters in a pre-merger phase has also been recently suggested by a few X-ray/optical observations (e.g. Murgia et al.2010; Maurogordato et al.2010).

\begin{figure} 
\includegraphics[width=0.495\textwidth]{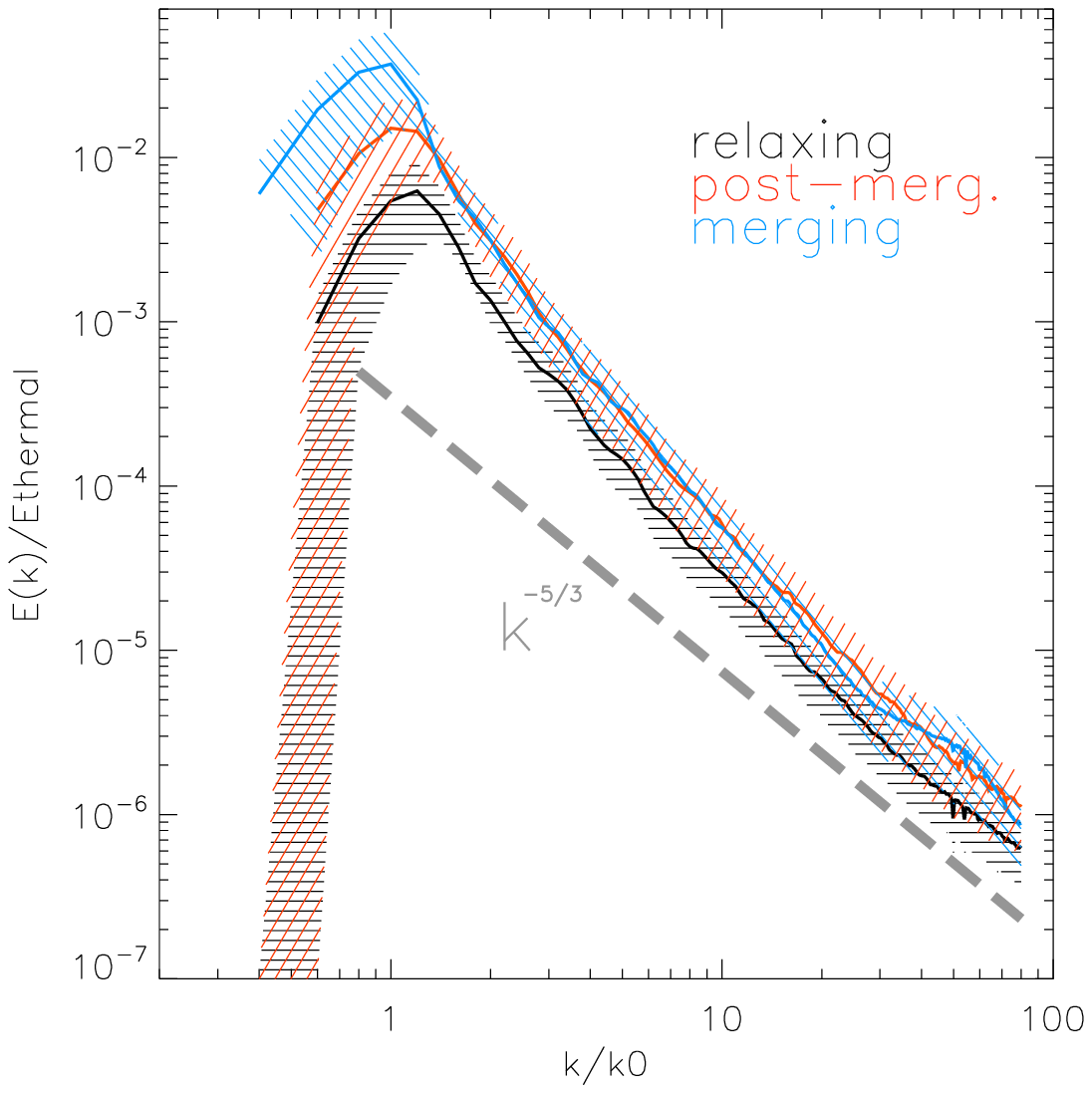}
\caption{Average power spectra of the 3--D velocity field for the 
different classes of galaxy clusters in our sample, at $z=0$.}
\label{fig:pk}
\end{figure}

\subsection{Spectral properties of the gas velocity fields.}
\label{subsec:spectra}

In Sec.\ref{subsec:turbo_method} we introduced the power spectrum of the 
3--D velocity field, $E(k)$, as useful tool to describe the spectral energy distribution of the ICM
in our simulations.

Figure \ref{fig:pk} shows the average
3--D power spectra for all clusters of the sample at $z=0$, after averaging each of the  
3 dynamical classes. The spatial frequency $k$ for every cluster has been referred to the one corresponding
to the virial radius, $k_{0} \approx 2\pi/R_{\mathrm vir}$, while $E(k)$ has been normalized
to the total thermal energy inside $R_{\mathrm vir}$.
The power spectra show a well-defined power law energy distribution
for nearly two orders of magnitude in spatial scale, with a slope of the order of  $\alpha \leq 5/3 - 2$ (where
$E(k) \sim k^{-\alpha}$). 
The maximum of $E(k)$ falls is in the range of  $\sim 1-2 R_{\mathrm vir}$ and the drop of the power
spectra at larger scales is clearly detected in our runs thanks to the large volume considered.
 This result basically confirms and extend to larger cluster masses the results obtained
 in Vazza et al.(2009) and Vazza, Gheller \& Brunetti (2010) and shows
 that the average spectral distribution of the 3--D velocity field in evolving galaxy
 clusters is self-similar across a wide range of virial masses. 
At spatial scales just below the maximum correlation scale, the slope of the power spectra gets significantly steeper compared to a standard $\alpha = 5/3$ slope of Kolmogorov turbulence, while
it becomes similar to the Kolmogorov behavior at 
intermediate spatial scales.
This may be explained as the effect of large scale bulk inflows through the cluster virial radius {\footnote{The motions are likely correlated on large scales due to geometrical reasons, see Vazza, Gheller \& Brunetti (2010).}}, whose 
kinetic energy adds to the kinetic energy carried by large scale turbulent motions. 
A qualitatively similar picture has been recently provided also by ENZO-MHD re-simulations of cluster formation (Xu et al.~2009,~2010).
It should also be noticed that the spatial scales smaller than $\leq 8 \Delta$  (where $\Delta$  is the cell size) are increasingly affected by  the numerical dissipation of the Parabolic Piecewise Method employed by ENZO 
(e.g.Porter \& Woodward 1994; Kitsionas et al.2009). However, the flattening of the spectral slope 
seen for $k/k_{0}>20$ is not dramatic.

\bigskip
 
The comparison of the three sub-samples shows that on average
merging clusters are characterized by significantly larger coherence
scales. This follows from the fact that, according to our definition (Paper I), we identify as merging systems those where a
close companion is found, in an early infalling phase. The larger
correlation at $\sim 1-2 R_{\mathrm vir}$ scales follows from
the presence of an early stage of interaction between the ICM of colliding
massive clusters (see also Fig.\ref{fig:turbo_mov}), whose virial volumes have just begun to cross each other at the moment of observation.

\bigskip

A complementary tool to measure the spectral features of the turbulent ICM is the
structure function of the 3--D velocity field. As in the case of the Inter Stellar
Medium, the structure functions (of various order) of the velocity field
provide a way to compare simulations with the theoretical expectations from the basic theory of turbulent flows (e.g.
Kritsuk et al.2007). In Vazza et al.(2009) we have shown that also in the case of
the simulated ICM the information provided by the transversal and longitudinal structure functions
give results consistent with the 3--D power spectra, showing maximum coherence scales of the
ICM flow at $\sim R_{\mathrm vir}$, and an overall good consistency with standard expectations from the Kolmogorov
scaling $S_{\mathrm 3}(\it l) \sim l$ (e.g. Kritsuk et al.2007).

Here we consider the 3rd order structure function of the absolute value of the velocity, $|{\bf v}|$: 

\begin{equation}
S_{\it 3}(\it l)=<|{\vec |v(r+l)|}-{\vec|v(r)|}|^{3}>.
\end{equation}

\begin{figure} 
\includegraphics[width=0.495\textwidth]{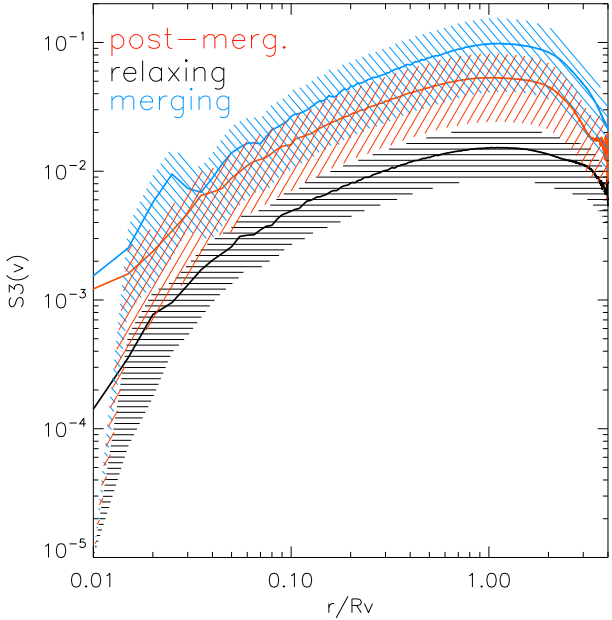}
\caption{3rd order structure functions for $|\vec {v(r+l)}-\vec{v(r)}|$ for the dynamical 
classes of the sample. The spatial scale has been normalized for the virial radius of every cluster, while $S_{\it 3}$ has been normalized by $c_{s}^{3}$ (where $c_{s}$ is the volume averaged sound speed within $R_{\mathrm vir}$).}
\label{fig:str3}
\end{figure}

For every cluster $S_{\it 3}$  was reconstructed 
by extracting $\sim 10^{6}$ random pairs of cells withing the AMR region and by 
measuring the average (volume-weighted) structure function in every
radial bin.
Figure \ref{fig:str3} shows the average results for the 3 dynamical classes of clusters. In this case the distance was rescaled to that
of the virial radius of every cluster, while the values of $S_{\it 3}$ have been
normalized to $c_{s}^{3}$  (where $c_{s}$ is the volume averaged sound speed inside
each cluster virial radius).
The maximum of the structure functions is found at the scale of $1-2 R_{\mathrm vir}$,
consistently with the outer scales provided by the power spectra
analysis. A  behavior broadly consistent with a linear scaling, $S_{\it 3}(\it l) \sim l^{0.5-1}$ is
found for all dynamical classes of clusters, within the range of $0.05 \leq r/R_{\mathrm vir} \leq 1.5$.

\begin{figure} 
\includegraphics[width=0.495\textwidth]{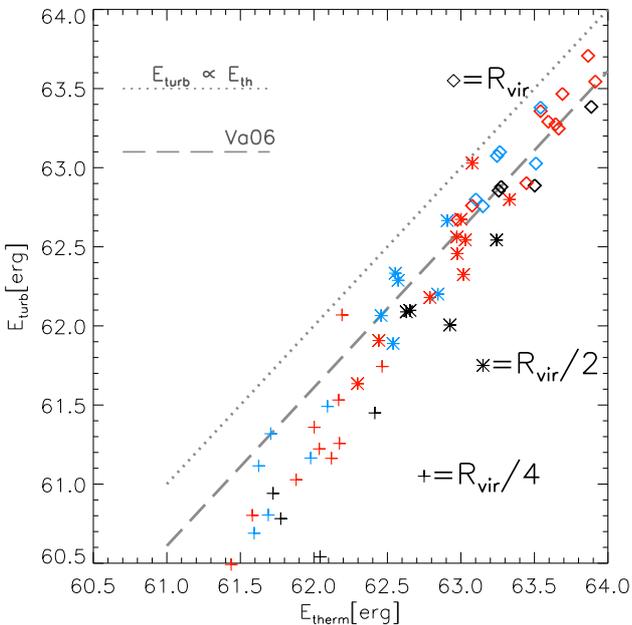}
\caption{Scaling between the turbulent energy and the
thermal energy inside a given radius, for three different radii: $<R_{\mathrm vir}$ 
(squares), $<0.5 R_{\mathrm vir}$ (stars) and $<0.25 R_{\mathrm vir}$ (crosses). The dotted
grey line shows the $E_{\mathrm turb} \propto E_{\mathrm therm}$ scaling, while the dashed gray line shows the 
$E_{\mathrm turb} \sim 0.27 E_{\mathrm therm}$ scaling reported in Vazza et al.(2006) for GADGET2 runs. 
The meaning of colors
for the data points is the same as in Fig.\ref{fig:prof_ratio}.} 
\label{fig:Eteth}
\end{figure}

\subsection{Scaling laws for the turbulent energy budget in clusters.}
\label{subsec:scalings}

The injection of chaotic motions at mergers/accretions is a mechanism 
mainly driven by the gravity of galaxy clusters, and therefore 
overall it should scale with virial cluster parameters. 
One way to model the turbulence injection in the ICM is to assume that during mergers
a fraction of the ${\it PdV}$ work done by the sub-clusters infalling onto the main cluster
goes into the excitation of turbulent motions (e.g. Cassano \& Brunetti 2005).
In this case turbulence  is injected in the cluster volume swept by the 
accreted sub-clusters, which is unbound by the effect of the ram pressure stripping.
Since the infalling sub-clusters are driven by the gravitational
potential, the velocity of the infall should be 
$\sim \sqrt{2}$ times the sound speed
of the main cluster. Consequently, the energy density of the turbulence injected during
the cluster--crossing should be proportional to the thermal energy density 
of the main cluster.
Also the fraction of the volume of the main
cluster in which turbulence is injected (the volume swept by the 
infalling sub-clusters) depends only on 
the mass ratio of the two merging clusters, provided that the distribution
of the accreted mass--fraction does not strongly depend on the cluster
mass (Lacey \& Cole 1993). The combination of these points implies that the
energy of turbulence should scale with the cluster thermal energy, 
$E_{\mathrm turb} \sim A \cdot E_{th} \sim M_{vir}^{3/2}$ (where
$A<1$, e.g. Cassano
\& Brunetti 2005). In agreement with these expectations, Vazza et al.(2006) derived best fit relations
for the scalings between turbulent energy, total virial mass and thermal energy in GADGET2 cosmological 
simulations, finding $E_{\mathrm turb} \propto M_{vir}^{1.6}$ and $E_{\mathrm turb} \sim 1/3 E_{th}$.

In Fig.\ref{fig:Eteth} we show the scaling between $E_{\mathrm turb}$ (measured
according to the "$k_{\mathrm MAX}$" filtering) and
$E_{\mathrm therm}$ for three different radii of integration: $r<0.25 R_{\mathrm vir}$, $r<0.5 R_{\mathrm vir}$
and $r< R_{\mathrm vir}$. In all cases the scaling closely follows 
$E_{\mathrm turb} \propto E_{\mathrm therm}$; the scatter however increases when 
smaller volumes are considered to compute the energies. 
This supports the idea that the turbulent energy measured in our simulations is a fully
gravitationally-driven process, for which the total mass is the driving parameters.
However variations of $E_ {\mathrm turb}/E_{\mathrm therm}$ are found at fixed masses
and in relation with the dynamical state of the cluster (see also Sec.\ref{subsec:time}).

The turbulent pressure support in the innermost regions of galaxy 
clusters is an important issue because all estimates of cluster
masses derived by X-ray observables are affected by
the presence of non-thermal pressure support at some level (Rasia 
et al.~2004; Lau et al.~2008; Piffaretti \& Valdarnini 2008; Burns, Skillman \& O'Shea 2010).  
Resolved spectra of the chaotic velocity field are beyond
the capabilities of existing X-ray facilities, at least for the amount of broadening 
seen by numerical simulations (e.g. Inogamov \& Sunyaev 2003; Dolag et al.~2005; Br\"{u}ggen, Hoeft \& Ruszkowski 2005; Vazza, Gheller \& Brunetti 2010). 
However, a number of upper limits
to the amount of turbulent motions in the central region of
galaxy clusters (e.g. within the cluster cores) are presently available (e.g. Churazov et al.~2008; Sanders et al.~2010a).
The comparison between these limits and our simulations provides a sanity check of our results.

Very recently, Sanders, Fabian \& Smith (2010) published upper-limits
to the turbulent velocity support available in the central region of 62 nearby galaxy clusters, galaxy groups and  elliptical galaxies. These authors find at least 15 cases with less than  $\sim 20$ per cent of the thermal 
energy density in turbulence, and weak evidence for turbulent velocities, of the order of $\sim 500~km ~s^{-1}$,
in the cool-core cluster RXJ1347.5-1145 that is undergoing a minor merger. It should be stressed that the results
are obtained after modeling and removing the contribution to the broadening of the spatial extent of the
objects, that is a difficult task (Sanders et al.2010b and discussion therein). Nowadays the most "direct" upper limit
is likely that obtained for A1835, which yields $E_{\mathrm turb}/E_{\mathrm therm} \leq 0.13$ on a scale of $\sim 30$~kpc
(Sanders et al.~2010a).

\begin{figure}
\includegraphics[width=0.495\textwidth, height=0.45\textwidth]{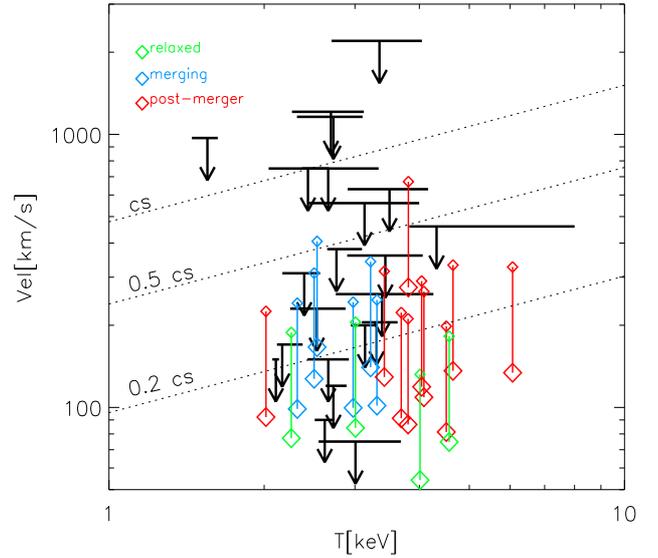}
\caption{Scaling between the average temperature and the mean velocity dispersion
(small squares) for all clusters of the simulated sample (squares in colors) and
for sample of the cluster observed with XMM-Newton by Sanders et al.(2010b). 
The additional dotted lines show the dependence of the ICM sound speed with the 
temperature. In order to compare with the observations, we filtered the velocity for the same  spatial coherence scale of $l \approx 30$ kpc available to Sanders et al.(2010b); the
data-point derived in this way are shown as connected thick squares.}
\label{fig:limits}
\end{figure}

Although upper limits to turbulence should be treated with caution, in Fig. \ref{fig:limits} we plot the upper limits derived from Sanders et al.~(2010b) for the velocity dispersion versus ICM temperature (as vertical arrows) of the 28 galaxy clusters in their sample, and the results from our simulations (as open squares).
The small top squares show the average value of the turbulent velocity field ("$k_{\mathrm MAX}$" method), while the connected
thick bottom squares show the turbulent energy associated with motion on  $\approx 30$~kpc scales (which roughly corresponds to the projected volume available to the observations of Sanders et al.~2010b).
Since the turbulent
energy spectrum at the smallest spatial scales in our simulations may be affected by numerical effect in the PPM code at the smallest scales
(e.g. Porter \& Woodward 1994 and Sec.\ref{subsec:spectra}), the energy of the turbulent motions on scales $ \leq 30$~kpc was derived analytically from the measured total power spectrum on larger
scales assuming Kolomogorov scaling. 

Figure \ref{fig:limits} shows that for most of our clusters the measured typical velocity dispersion of motions expected at scales $l<30$~kpc
is of the order of $\sigma_{v} \sim 0.2c_{s}$ or smaller,  well below the upper limits obtained by {\it XMM-Newton}.
It is interesting to notice that, even in the case of merging
and post-merger systems, where usually very large ($l \sim $Mpc) chaotic eddies with $E_{\mathrm turb}/E_{\mathrm therm} \sim 0.2-0.3$ can develop, the turbulent energy available to the smallest scales is 
consistent with the upper limits of Sanders et al.~(2010b). 
We should however notice that  additional turbulent injection
mechanisms connected with AGN outflows are expected to increase the amount of turbulent mixing
around cool core clusters (e.g. Br\"uggen et al. 2005; Heinz et al. 2010; Dubois et al.~2010) and that
 therefore our estimates here can only be strictly compared with non-cool core clusters.
\noindent  Importantly, as the authors noted, the upper limits reported by Sanders et al.(2010b) may be subject to large uncertainties
 due to the rather uncertain modeling of the thermal gas distribution around the target sources. 
Thanks to the high spectral resolution, in a few years {\it Astro-H} will likely provide substantial step in our
understanding of turbulent motions in the ICM (e.g. Takahashi et al.~2010).

\begin{figure} 
\includegraphics[width=0.45\textwidth]{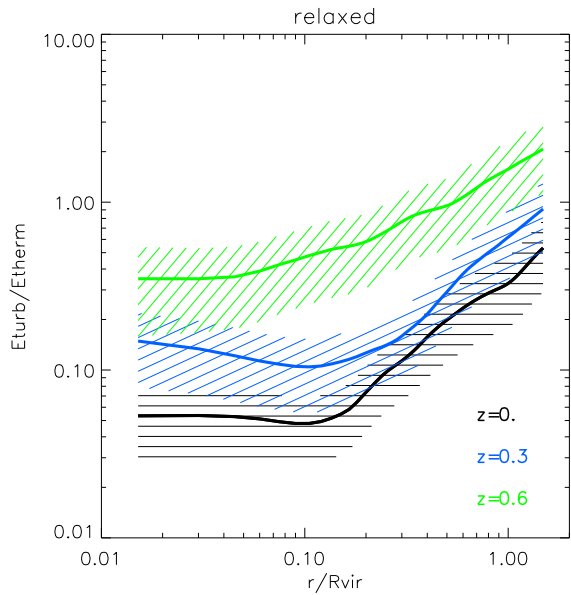}
\caption{Radial profiles of the $E_{\mathrm turb}/E_{\mathrm therm}$ ratio for the clusters classified as relaxed 
at $z=0$.}
\label{fig:prof_evol}
\end{figure}

\begin{figure} 
\includegraphics[width=0.495\textwidth]{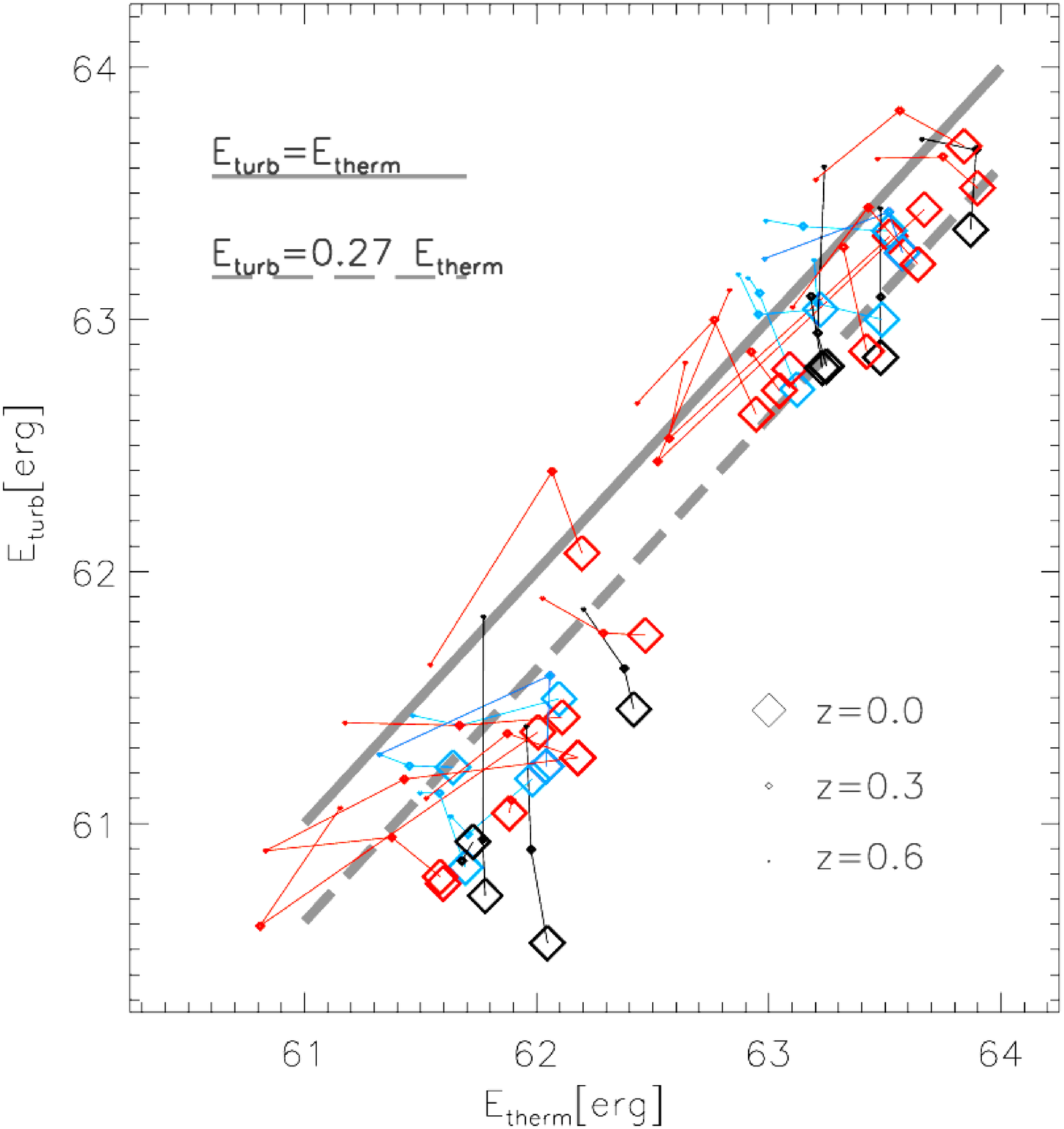}
\caption{Scaling between the turbulent energy and the
thermal energy inside a given radius, as estimated for $r<R_{\mathrm vir}$ (top squares) and for $r<0.25 R_{\mathrm vir}$ (lower squares). The dotted
gray line shows the $E_{\mathrm turb}=E_{\mathrm therm}$ scaling, while the long dashed
one shows the $E_{\mathrm turb}=0.27 E_{\mathrm therm}$ best fit relation of Vazza et al.(2006). The meaning of colors
is the same as in Fig.\ref{fig:prof_ratio}.} 
\label{fig:Eteth_z}
\end{figure}

\subsection{Time evolution of turbulence.}
\label{subsec:time}

The evolution of the total mass for each cluster of our
sample has been presented in Paper I (Fig.4). 
These results have shown that the bulk of the total (gas+DM) mass of half of 
our clusters is built in major merger events at $z<1$.
This implies that only a few dynamical times have past since 
the major merger forming our post-merger clusters.  Therefore, correlations
between turbulent energy and the epoch of the last violent merger should be expected.

In order to monitor time-dependent behaviors of turbulence in our clusters sample, we measured
the turbulent energy (with the $k_{\mathrm MAX}$ method) and the thermal energy also for the additional redshifts of  $z=0.6$ (look back time of $\approx 5.6$ Gyr) and $z=0.3$ ($\approx 3.3$ Gyr). 
Figure \ref{fig:prof_evol} illustrates the evolution of the profile of $E_{\mathrm turb}/E_{\mathrm therm}$ 
for the clusters classified as relaxed at $z=0$; the averaging procedure is the same of Sec.\ref{subsec:scalings}.
These systems present more
chaotic motions at all radii at increasing redshifts; in particular
at the reference epoch of $z=0.6$ their normalized profile is on average rather similar to that of major merger
systems at $z=0$, suggesting that the ratio $E_{\mathrm turb}/E_{\mathrm therm}$ is strictly dependent on the look-back time since the last major merger.

In Fig.\ref{fig:Eteth_z} we show the values $E_{\mathrm turb}/E_{\mathrm therm}$ integrated inside spheres of $r<0.25 R_{\mathrm vir}$
ande $r<R_{\mathrm vir}$ for $z=0.3$ and $z=0.6$ for all the clusters in our sample (using the "$k_{\mathrm MAX}$" method). 
Also in this case the time evolution is rather different when the 3 dynamical
classes are compared.
Figure \ref{fig:Eteth_z} demonstrates the transient feature of turbulence in our simulated
clusters: variations of $E_{\mathrm turb}/E_{\mathrm therm} \sim 10$ are observed in a few Giga-years in connection
with the dynamical status of clusters. 
Relaxed systems move preferentially "downwards" in the ($E_{\mathrm therm},E_{\mathrm turb}$)   
plane (i.e. by dissipating their turbulent
energy and slightly increasing their thermal energy with time), post-merger systems often present a peak of
turbulent injection at intermediate redshifts (due to the merger event they all experience in the $0 \leq z \leq 1$ range), while merging systems present slightly more irregular paths across the same plane.
A very similar behavior is found also by adopting the filter $l_{\mathrm MAX}=300$kpc (see also Vazza et al.2006).

\begin{figure} 
\includegraphics[width=0.49\textwidth]{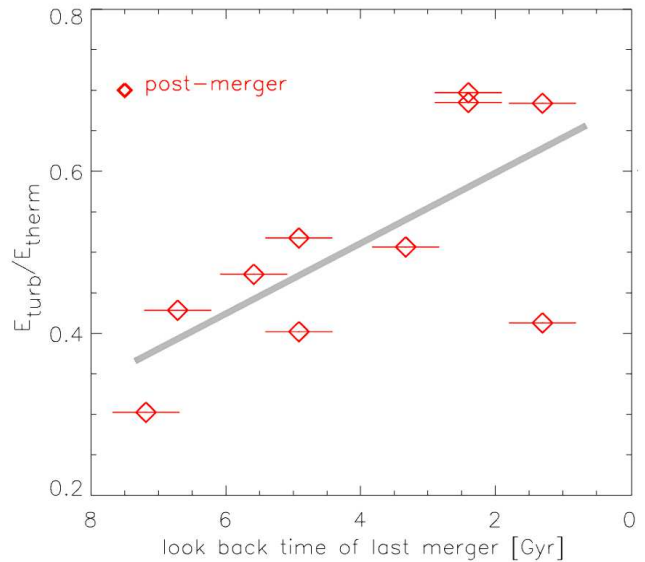}
\caption{Energy ratio $E_{\mathrm turb}/E_{thermal}$ at $z=0$ versus the look-back time of the last merger
event for the post-merger systems of the datasample; the horizontal errorbars show the approximate
uncertainty in the exact epoch of the central phase of the major merger event for each cluster. }
\label{fig:Et_time}
\end{figure}

Figure \ref{fig:Et_time} shows the relation between $E_{\mathrm turb}/E_{\mathrm therm}$ at $z=0$ 
(inside $r<R_{\mathrm vir}$) and 
the look-back time from the last major merger, $t_{LM}$, for the post-merger systems (adopting the "$l_{\mathrm MAX}$'' method).
For these clusters we can identify the epoch of the last major merger with an uncertainty of the order of a cluster-cluster crossing time ($\sim 1-2$~Gyr).
This Figure supports a simple trend between $E_{\mathrm turb}/E_{\mathrm therm}$ and $t_{LM}$,
of the type  $E_{\mathrm turb}/E_{\mathrm therm} \propto - \beta t_{LM}$, with $\beta \approx 2/3$, and provides additional support to the merger-turbulence connection
in our simulated clusters.

\bigskip

\begin{figure*}
\begin{center}
\includegraphics[width=0.45\textwidth]{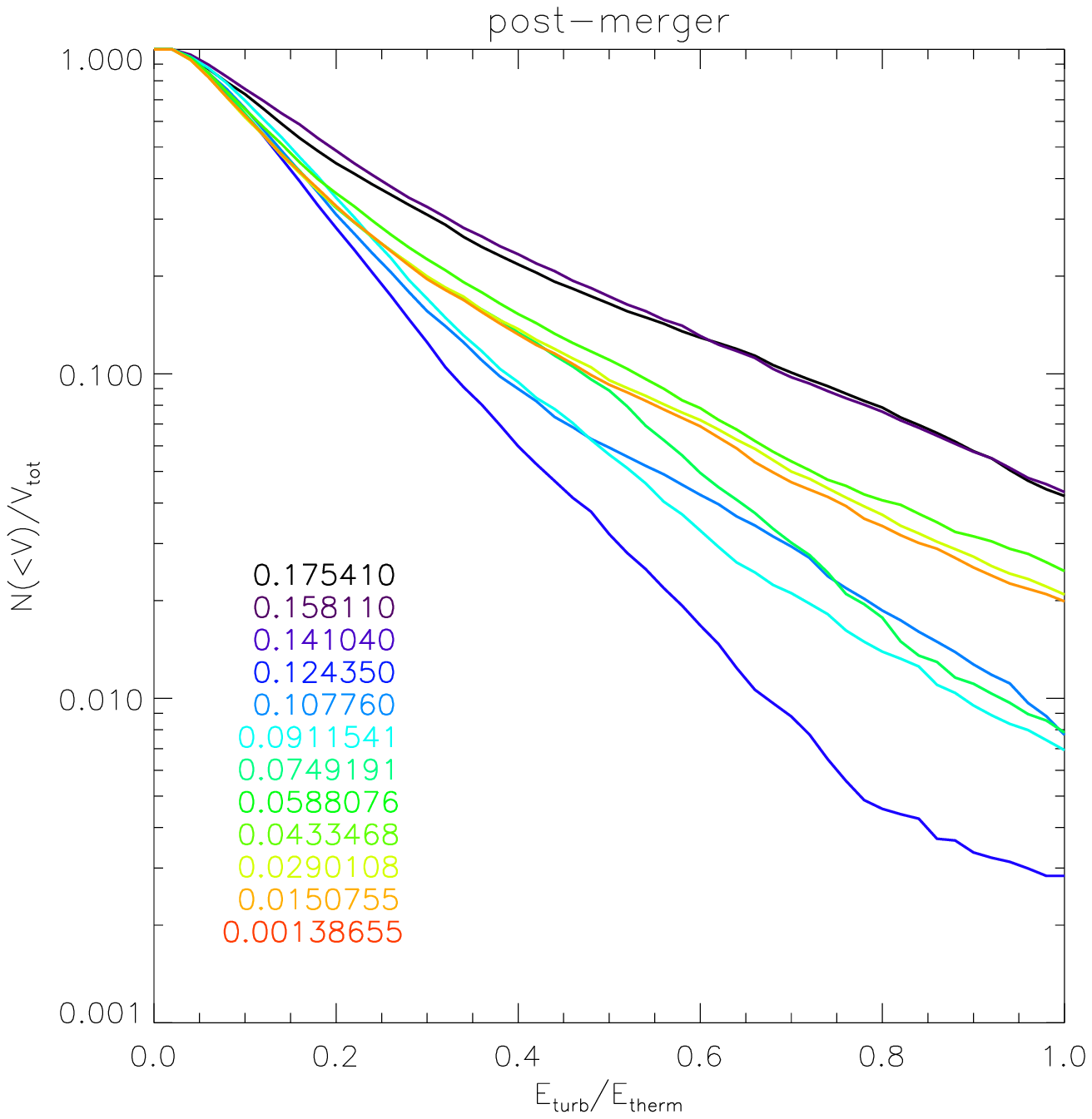}
\includegraphics[width=0.45\textwidth]{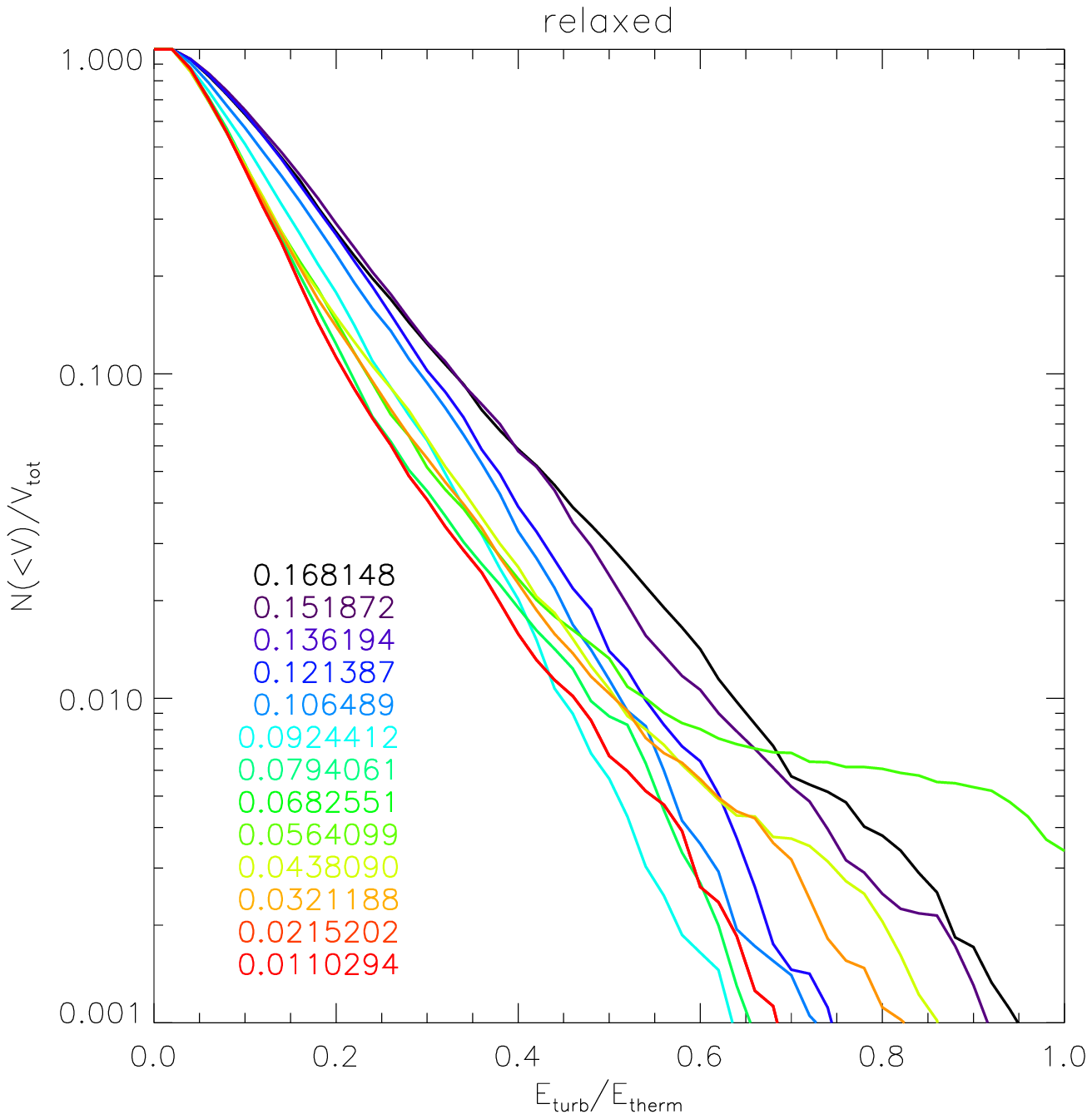}
\caption{Cumulative volume distribution functions for the ratio between turbulent energy
and thermal energy inside $\sim 1Mpc^{3}$ for two clusters in the sample,
by adopting the filtering with $l_{\mathrm MAX}=300$kpc.}
\label{fig:dist_turb}
\end{center}
\end{figure*}

\subsection{On the connection between turbulence and giant radio halos}
\label{subsec:halos}
 
Radio halos are Mpc synchrotron radio sources detected in the central regions of a fraction of 
galaxy clusters (Feretti \& Giovannini 2000; Kempner \& Sarazin 2001; Van Weeren et al.~2009; Ferrari et al.~2008 for a modern review). These emissions are found preferentially in X-ray luminous galaxy clusters (e.g. Giovannini, Tordi \& Feretti 1999; Venturi et al.~2008; Giovannini et al.~2009) and only in systems with a X-ray perturbed morphology, showing indications of merger activity (Buote 2001). Nowadays clear evidence of a statistical
connection between cluster  mergers 
and radio halos is provided by present data (Cassano et al.2010).
In principle, several  physical mechanisms may contribute to the origin of non thermal 
components to power such large scale radio emission
(e.g. Blasi \& Colafrancesco 1999; Sarazin 1999; Dolag \& Ensslin 2000; Brunetti et al.~2001)

Turbulence in the magnetized ICM may power large
scale radio emission in galaxy clusters, provided that relativistic electrons 
can couple with the MHD modes excited during merger events (e.g. Brunetti et al.~2008
and references therein).
Brunetti \& Lazarian (2007, 2010) developed a comprehensive modeling of the properties of MHD
turbulence in the ICM and of turbulent acceleration of relativistic particles. According to their
picture compressible turbulence provides the most important contribution to the process of 
particle (turbulent) re-acceleration in the ICM. Under the assumption of a collisional coupling
between the turbulent modes and both the thermal and relativistic particles in the ICM, these
authors have shown that radio halos can be switched on in massive clusters when compressible
turbulence (generated on $\sim 300$~kpc scales) accounts for $\geq 15-30$ per cent of the thermal
energy. 
Although we have shown in Sec.\ref{subsec:radial}
that in merging and post-merger clusters $E_{\mathrm turb}/E_{\mathrm therm}$ is fairly large, turbulence should also be rather volume filling in order to produce radio halos of $\sim$~Mpc size. 

In Fig.\ref{fig:dist_turb} we show the low redshift ($z<0.2$) evolution of the cumulative volume distribution of
$E_{\mathrm turb}/E_{\mathrm therm}$,  inside a reference fixed
volume of the order of that filled by typical radio halo ($\approx 1 Mpc^{3}$) centered in the cluster center of two massive
systems of our sample, with nearly identical final mass: a relaxed cluster at $z=0$ (right) and a 
major merger cluster (left). 
To be conservative in Fig.\ref{fig:dist_turb} we used the $l_{MAX}=300$~kpc filter, which likely provides a lower limit
to turbulent motions in the ICM. 

The fraction of volume where  $E_{\mathrm turb}/E_{\mathrm therm}$ is above a fixed threshold is larger in the post-merger cluster. At $z \sim 0 $ only $\sim 5$ per cent of this central volume  has
$E_{\mathrm turb}/E_{\mathrm therm} >0.3$ in the relaxed cluster, compared to a $\sim 30-40$ per cent in the post-merger cluster. 
Although turbulence within the scale length of $<300$~kpc is not totally volume filling even in the post-merger cluster, the whole total "projected" volume within the central $1~Mpc^{3}$ of the cluster would be seen as "turbulent" from every line of sight (e.g. Subramanian
et al.~2006; Iapichino \& Niemeyer 2008). 
Similar results can be found for all the clusters within the sample, showing a very tight correlation between the volume filling factor of turbulence and the dynamical state of the host galaxy cluster.

\bigskip

\begin{figure}
\includegraphics[width=0.45\textwidth,height=0.4\textwidth]{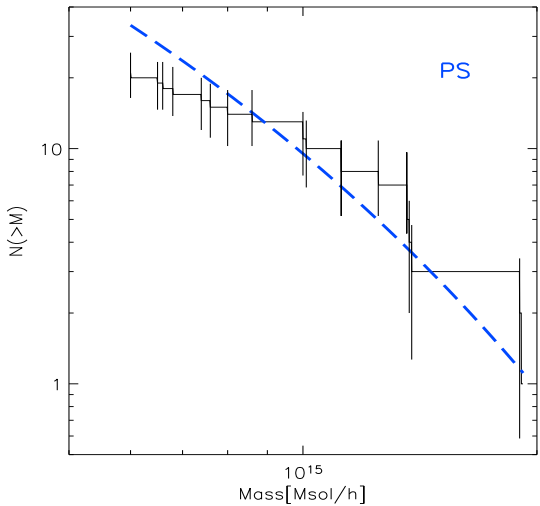}
\caption{Cumulative mass function for the clusters of our sample (solid line)
compared to the theoretical Press-Schecter mass function for the same volume (dashed line). The  vertical error bars mark the Poissonian noise for every mass bin. }
\label{fig:mass_funct}
\end{figure}

Our sample is large enough to allow a first statistical estimate of the frequency of "turbulent" clusters and
a first comparison with the observed frequency of radio halos
in real galaxy clusters. The cluster-mass distribution of  completeness of our sample is  consistent with the Press \& Schecter mass function 
for $M \geq 7 \cdot 10^{14}M_{\sun}/h$; for lower masses, the lack of objects
is of the order of $\sim 30$ per cent (Fig.\ref{fig:mass_funct}).
This shows that our sample can be safely used for statistical
studies provided that our clusters are not biased with respect to a particular dynamical state (as in
our case, see Paper I).

\begin{figure*} 
\includegraphics[width=0.45\textwidth]{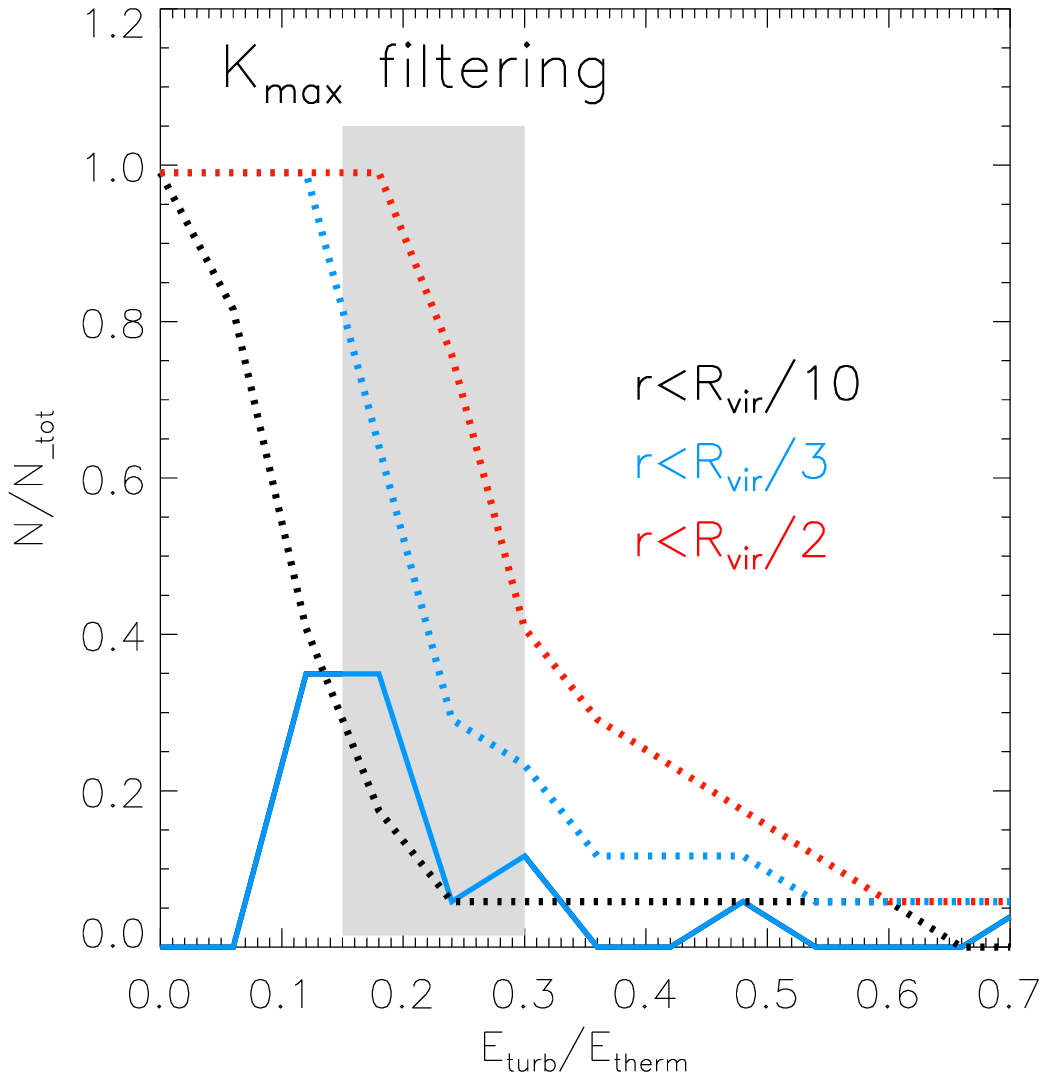}
\includegraphics[width=0.45\textwidth]{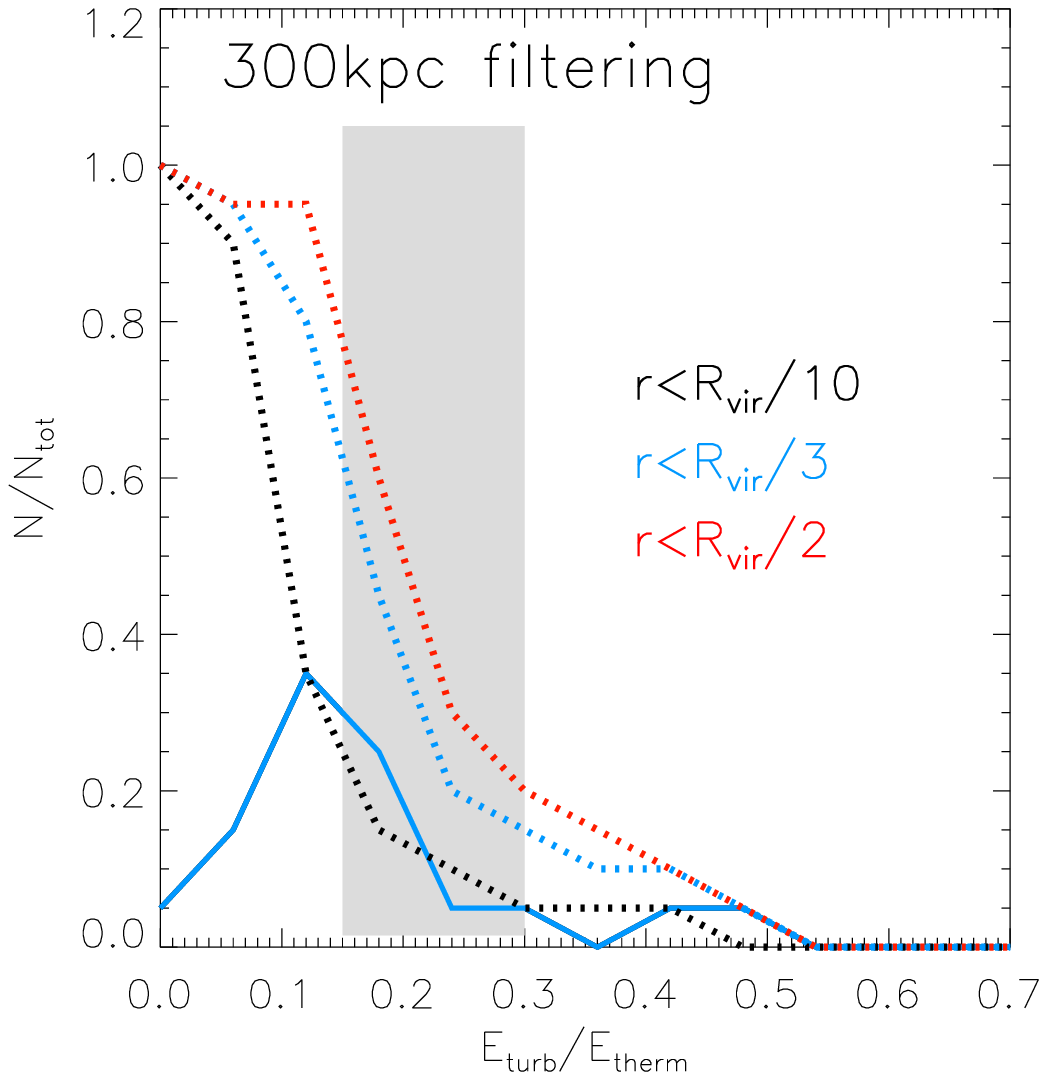}
\caption{Distribution functions for $E_{\mathrm turb}/E_{\mathrm therm}$ inside three
reference radii ($R_{\mathrm vir}/2$, $R_{\mathrm vir}/3$  and $R_{\mathrm vir}/10$) for
the simulated clusters at $z=0$. The left panel is for the turbulent
ICM velocity field estimated using the $l_{\mathrm MAX}$ filtering, while the right
panel is for the filtering at $l_{\mathrm MAX}=300$kpc. The solid lines show the differential distributions (for $r<R_{\mathrm vir}/3$ only), while the dashed lines show the cumulative distributions. The vertical grey band shows the approximate regime of turbulence required by present calculation ofr the turbulent re-acceleration of relativistic electrons (e.g. Brunetti \& Lazarian 2007).}
\label{fig:halos_freq}
\end{figure*}

According to calculations of turbulent acceleration in galaxy clusters an energy ratio of $E_{\mathrm turb}/E_{\mathrm therm} \approx 0.2 \cdot \sqrt{5/(k_{B}T_{\mathrm keV})}$ (where $T_{\mathrm keV}$ is the cluster temperature measured in $keV$) may allow the generation of radio halos (Brunetti \& Lazarian 2007) {\footnote{By considering compressible turbulence
generated at $\sim 300$~kpc scale.}}.
In Fig.\ref{fig:halos_freq} we show the distribution of $E_{\mathrm turb}/E_{\mathrm therm}$ for the clusters in our sample,  evaluated in the typical
volume occupied by radio halos.
The size of observed radio halos is usually in the range of
$\sim Mpc$, that corresponds to  $\sim R_{\mathrm vir}/3$
for most of the clusters of our sample (for completeness we also report the distribution for
$r<R_{\mathrm vir}/10$ and $R_{\mathrm vir}/2$ in the same Figure).
The cumulative distributions  extracted for $<R_{\mathrm vir}/2$ and $R_{\mathrm vir}/3$ are similar for the $l_{\mathrm MAX}=300$kpc filtering (right panel), while the turbulent
energy ratio increases in the $l_{\mathrm MAX}$ filtering when larger volumes are considered for the integration; this may
be explained with the large scale bulk motions present outside of the cluster innermost region.
 Regardless of the filtering technique, 
 $\sim 1/3$ of the massive galaxy clusters in the local Universe is found to host
turbulent motions of the order of $E_{\mathrm therm}/4$ within $R_{\mathrm vir}/3$.
This suggests that the theoretical conditions for the generation of radio halos in our simulated clusters
are achieved for $\sim 1/3$ of our clusters, provided that turbulence drives the formation
of diffuse radio emission.
Nowadays extensive radio follow ups of complete samples of galaxy clusters indicate that a $\sim 1/3$ of massive
$L_{\mathrm X} > 5 \cdot 10^{44} erg/s$ ($M>10^{15}M_{\sun}$) objects host Mpc-scale halos (e.g. Venturi 
et al.2007; 08; Cassano et al.2008) and that all radio halo clusters are merging systems (e.g. Cassano et al.2010).
This is consistent with our previous claim.
Semi-analytical calculations based on extended Press-Schechter theory already demonstrated that the theoretical
occurrence of Mpc-scale radio halos (assuming that they originate from turbulent acceleration) is consistent with the
occurrence of these sources as measured by present surveys (Cassano et al.2008).
This is the first time in which an overall consistency between the fraction of "turbulent" clusters and that
of radio-halos clusters in massive objects is underlined by means of numerical cosmological simulations.

\section{Discussion and conclusions}
\label{sec:conclusions}

In this paper we explore the properties of turbulent motions in a statistical sample
of massive galaxy clusters simulated at high-resolution with the AMR code ENZO (Norman et al.2007).

We focus on a sample of 20 massive galaxy clusters already presented in Paper I and simulated
with the tailored AMR technique introduced in Vazza et al.(2009). We provide a complete view of the onset and of the evolution of gravity-driven turbulent motions which are  injected in massive galaxy clusters via mergers.

However, the physical modeling adopted in our
simulations does not include the effect of AGN activity, Cosmic Rays and magnetic field.
In reality, AGN activity may drive small-scale turbulent motions around the core region
of clusters (e.g. Heinz et al.2006; Scannapieco \& Bruggen 2009; Dubois et al.2010, Morsony et al.2010; Vazza 2010), while the interplay of tangled magnetic field and Cosmic Rays is expected 
to inject chaotic motions at very small scales, via instabilites in the ICM (Parrish
\& Stone 2008; Quataert 2008; Ruskowsky \& Oh 2010; McCourt et al.2010; Ruszkowsky et al.2010).

The magnetic field is also expected to play a role along the cluster evolution (e.g.  Dolag~2006; Dubois \& Teyssier 2008;Dolag \& Stasyszyn~2009; Donnert et al.~2009;  Collins et al.~2010), even if without affecting the spectral properties outlined in Sec.\ref{subsec:spectra} for a wide range of scales (e.g. Xu et al.2009; Xu et al.2010). 

The implicit assumption to trust in the development of the turbulent motions as derived from our simulations
is that kinematic
viscosity in the ICM is negligible for all scales larger than the minimum
cell size, $25$~$kpc ~h^{-1}$, which implies that the effective kinematic
viscosity, $\nu$, must be smaller than $\nu \sim 10^{29} {\mathrm cm}^{2} {\mathrm ~s^{-1}}$
{\footnote{This can be roughly estimated considering that at the scale
of the minimum cell size, $\Delta x=25$~kpc, the typical value of the dispersion
in the velocity field is $v < 50$~$km ~s^{-1}$, as shown in Fig.\ref{fig:limits},
and the Reynolds number is equal to one, and therefore : 
$\nu \sim 50{\mathrm km ~s^{-1}} \cdot 25 {\mathrm ~ kpc ~h^{-1}} / R_{\mathrm e} \sim 10^{29}{\mathrm cm}^{2} {\mathrm ~s^{-1}}$.}}.
At smaller scales there are physical reasons to believe that the turbulence in the ICM is finally dissipated in a collision-less
regime, e.g. by accelerating relativistic particles (Brunetti \& Lazarian 2007). 
A correct modeling of the dynamics of turbulent motions at the scales 
between the numerical resolution of the scheme and the physical scale responsible for dissipation (which is likely $<<kpc$)
can be computed by means of sub-grid modeling, by assuming some closure
relation for the turbulent equations. Valuable attempts have been
recently done in the framework of galaxies and galaxy clusters simulations
(e.g. Scannapieco \& Bruggen 2008; Maier et al.2009; Iapichino et al.2010); however the
application of these techniques in astrophysics is still in its beginnings.

\bigskip

To summarize our results, the analysis of turbulent motions in this sample of massive
galaxy clusters extends the previous results based on the same AMR technique
in ENZO and focused on clusters with smaller masses (Vazza et al.2009; Vazza, Gheller \& Brunetti 2010). 
Given the fairly large number of objects in this present sample (20) we were able to study the dependence of the turbulent
features on the dynamical state of clusters, following from their matter accretion history across cosmic time.
Two methods were presented to detect turbulent motions in the ICM. One is based
on a filtering in the Fourier space of the component of velocities associated with wave numbers larger than
the wavenumber of the maximum spectral energy, and the other is based on the filtering in the real space
of the velocity component with coherence scales smaller than the fixed length of $l_{\mathrm MAX}=300$~kpc.
(Sec.\ref{subsec:turbo_method}).
We have shown that the statistical results obtained in this paper are largely independent of the particular
method adopted. 
Post-merger and merging clusters show large values of turbulent energy compared to the
thermal energy of the ICM, with $E_{\mathrm turb}/E_{\mathrm therm} \sim 0.2-0.3$ in the innermost cluster regions.
On the other hand relaxed
clusters show much lower values of the turbulent ratio, $E_{\mathrm turb}/E_{\mathrm therm} \sim 0.05$ (Sec.\ref{subsec:radial}) within the same radius. These results are in line with recent studies performed
by Paul et al.~(2010), but extends to clusters with larger masses. 

Owing to the very high dynamical range achieved in our simulations ($N \sim 550$) we were able
to study the spectral features of the ICM velocity field with an unprecedented separation between the forcing and the dissipation scales, achieving a typical Reynolds number of the order of $R_{\mathrm e} \sim 500-1000$.
The power spectra of the 3--D velocity fields extend across nearly two 
orders of magnitude, with $E(k) \sim k^{-5/3 \div -2}$, and  show a typical scale for the peak of the energy 
spectrum at the scales of $1-2 R_{\mathrm vir}$ (with the tendency of merging clusters to present the largest
outer correlation scales). Consistent results were also obtained by the measure of the third-order structure function
of the velocity field across the clusters sample (Sec.\ref{subsec:spectra}).
As a sanity check we compared our results with the available limits on turbulent motions in the ICM for X-ray
observations (Sanders et al.~2010b).
We show that the available limits on turbulent motions on $\sim 10-50$~kpc scales obtained in the compact cores of 
relaxed clusters, $E_{\mathrm turb}/E_{\mathrm therm} < 10$ per cent, are well consistent
with the amount of these motions measured in our simulated clusters (Sec.\ref{subsec:scalings}). 
We derive scaling laws for the integrated turbulent energy in our clusters,  and confirm 
previous results based on GADGET2 runs (Dolag et al.~2005; Vazza et al.~2006), where
turbulence is found to scale with the thermal energies (Sec.\ref{subsec:scalings}). 
The time evolution of the turbulent energy
within clusters was sampled at different epochs ($z=0$, $z=0.3$ and $z=0.6$), reporting a significant
similarity between 
the turbulence found in the past of relaxed clusters and that found in merging clusters at $z=0$ (Sec.\ref{subsec:time}). We also found an anti-correlation between the look-back time since the
last major merger of a cluster and the level of turbulence at $z=0$.

Finally we explore the connection between turbulence and radio halos in galaxy clusters. Thanks to the
unprecedented statistics of our sample of massive clusters simulated with AMR, we can compare the
occurrence of turbulence in our clusters  with that of observed giant radio halos in nearby massive clusters.
Current calculations of turbulent acceleration for the origin of radio halos suggest that these sources can be generated in the turbulent ICM (e.g. Brunetti \& Lazarian 2007). 
We found that these conditions are reached only in the case of merging systems, where $\sim 1/3$ of the cluster 
volume is in the form of turbulent motions for a few Gyr, while only a few percent of the cluster
volume is turbulent in the case of relaxed systems (Sec.\ref{subsec:halos}).
In particular we find that $\sim 1/3$ of our clusters show $E_{\mathrm turb}/E_{\mathrm therm} > 0.15-0.30$ level
on Mpc-scale and we notice that this fraction is consistent with that of massive clusters hosting giant
radio halos.
These findings are in line with previous results based on extended Press-Schechter theory, that show that
the theoretical occurrence of radio halos in the turbulent acceleration model is consistent with present
observations.

The generation and evolution of turbulent motions in our simulated ICM  is very complex, and it  
follows from the multi scale energy distribution of the turbulent ICM, as produced by the dynamical evolution of galaxy clusters. 
It is very intriguing that our rather simple
physical setup  (which does not consider any non-gravitational mechanism) is sufficient 
to capture some of the most important multi scale features at work in the turbulent ICM and to
produce results in line with existing observations.
The adoption of an ad-hoc Adaptive Mesh Refinement scheme specially 
tailored is mandatory to  capture the full range of scales needed to describe the evolution of turbulence in the 
ICM; however in the future more sophisticated physical modeling including magnetic fields,Cosmic Rays would
be needed to model also the morphological and spectral features of synchrotron radio emission in a detailed
and time-dependent way.

 \section{Acknowledgments}
 
 We acknowledge partial 
support through grant ASI-INAF I/088/06/0 and PRIN INAF 2007/2008, and the 
usage of computational resources under the CINECA-INAF 2008-2010 agreement
and the 2009 Key Project ``Turbulence, shocks and cosmic rays electrons 
in massive galaxy clusters at high resolution''. F.V. and M.B. acknowledge support
from the grant FOR1254 from Deutschen Forschungsgemeinschaft. 
We thank J. Sanders for
kindly making us available his data points for Fig.\ref{fig:limits}.
We acknowledge S.Ettori, R.Valdarnini, J. Donnert, A.Kritsuk and A.Bonafede of very fruitful discussions.
We thank the anonymous referee for his useful comments to this paper.
We thank A. J. Br\"{u}ggen for his late, but fundamental contribution to our work.

\bigskip
\end{document}